\documentclass[fleqn,usenatbib]{mnras}

\usepackage{newtxtext,newtxmath}

\usepackage[T1]{fontenc}

\DeclareRobustCommand{\VAN}[3]{#2}
\let\VANthebibliography\thebibliography
\def\thebibliography{\DeclareRobustCommand{\VAN}[3]{##3}\VANthebibliography}

\usepackage{graphicx}	
\usepackage{amsmath}	
\usepackage{orcidlink}
\usepackage{xcolor} 
\usepackage{ulem} 

\title[Stellar-mass  Black Hole TDEs]{\vspace{-0.75cm} Wind-Reprocessed Transients from Stellar-Mass Black Hole Tidal Disruption Events}

\author[Kremer et al.]{
\parbox[t]{\textwidth}{\vspace{-0.75cm}
Kyle Kremer\orcidlink{0000-0002-4086-3180}
,$^{1,2}$\thanks{E-mail: kkremer@caltech.edu}\thanks{NASA Einstein Fellow}
Brenna Mockler\orcidlink{0000-0001-6350-8168}
,$^{2}$
Anthony L. Piro\orcidlink{0000-0001-6806-0673}
$^{2}$
and James C.\ Lombardi, Jr.\orcidlink{0000-0002-7444-7599}
$^{3}$} \vspace{-0.25cm} \\
$^{1}$ TAPIR, California Institute of Technology, Pasadena, CA 91125, USA\\
$^{2}$The Observatories of the Carnegie Institution for Science, Pasadena, CA 91101, USA\\
$^{3}$Department of Physics, Allegheny College, Meadville, Pennsylvania 16335, USA
\vspace{-0.5cm}
}

\date{\vspace{-0.5cm}\today}
\pubyear{2023}
\begin{document}
\maketitle

\begin{abstract}
Tidal disruptions of stars by stellar-mass black holes are expected to occur frequently in dense star clusters. Building upon previous studies that performed hydrodynamic simulations of these encounters, we explore the formation and long-term evolution of the thick, super-Eddington accretion disks formed. We build a disk model that includes fallback of material from the tidal disruption, accretion onto the black hole, and disk mass losses through winds launched in association with the super-Eddington flow. We demonstrate that bright transients are expected when radiation from the central engine powered by accretion onto the black hole is reprocessed at large radii by the optically-thick disk wind. By combining hydrodynamic simulations of these disruption events with our disk+wind model, we compute light curves of these wind-reprocessed transients for a wide range of stellar masses and encounter penetration depths. We find typical peak bolometric luminosities of roughly $10^{41}-10^{44}\,$erg/s (depending mostly on accretion physics parameters) and temperatures of roughly $10^5-10^6\,$K, suggesting peak emission in the ultraviolet/blue bands. We predict all-sky surveys such as the Vera Rubin Observatory and ULTRASAT will detect up to thousands of these events per year in dense star clusters out to distances of several Gpc.
\end{abstract}

\begin{keywords} transients: tidal disruption events --
stars: black holes -- globular clusters: general -- hydrodynamics -- accretion discs \vspace{-0.75cm}
\end{keywords}

\section{Introduction}

The presence of stellar-mass black hole populations in dense star clusters has gained considerable interest in recent years. For a cluster with $N$ stars following a standard initial stellar mass function \citep[e.g.,][]{Kroupa2001}, it is nearly certain that a subset of sufficiently massive stars will collapse into black holes on timescales $\lesssim50\,$Myr. Less certain are the prospects for retaining these black holes throughout the subsequent evolution of the host cluster, and specifically to the present day. Natal kicks \citep[e.g.,][]{Repetto2012}, gravitational dynamics \citep[e.g.,][]{Spitzer1969}, and recoil kicks associated with gravitational-wave-driven mergers \citep[e.g.,][]{Lousto2010,GerosaKesden2016} all act to eject stellar-mass black holes from their host cluster. Indeed, for clusters comparable to or lower in mass relative to the Milky Way globular clusters ($M_{\rm cl} \lesssim 10^6\,M_{\odot}$), it was thought for many years that the cumulative effect of these ejection mechanisms would prevent the long-term retention (beyond a few Gyr) of all but a handful of black holes \citep[e.g.,][]{Kulkarni1993}.

However, in the past decade observational evidence of stellar-mass black hole binaries in a number of Milky Way globular clusters through both dynamical measurements \citep[e.g.,][]{Giesers2018,Giesers2019} and X-ray/radio measurements \citep[e.g.,][]{Strader2012,MillerJones2015} have demonstrated at least some globular clusters can retain their black holes to the present day. These observations have been complemented by state-of-the-art $N$-body simulations \citep[e.g.,][]{Morscher2015,wang2016dragon,ArcaSedda2018,kremer2020modeling} which demonstrate that \citep[pending uncertainties regarding cluster initial conditions; e.g.,][]{PortegiesZwart2000,kremer2019initial}, a significant number (tens to hundreds) of black holes are expected to be retained to the present day in most globular clusters \citep[e.g.,][]{Weatherford2020}.

The presence of these stellar-mass black holes in dense star clusters leads naturally to a number of implications. For one, black holes can dynamically exchange into binaries with stellar companions \citep[e.g.,][]{Kremer2018_xrb}. In addition to forming the aforementioned in-cluster sources, these black hole-star binaries can also be dynamically ejected from their host cluster \citep[e.g.,][]{Giesler2018}, potentially providing a formation mechanism for the growing number of detached black hole binaries observed in the Galactic field by \textit{Gaia} \citep[e.g.,][]{ElBadry2023a}, whose formation is difficult to explain through standard isolated binary evolution scenarios. Secondly, black holes in clusters naturally pair up with other black holes eventually leading to binary black hole mergers \citep[e.g.,][]{PortegiesZwart2000,Rodriguez2016} plausibly similar to those detected by LIGO/Virgo as gravitational wave sources \citep[e.g.,][]{LIGO2016,LIGO2021}. Recent studies suggest a potentially large fraction of the LIGO black hole mergers may have originated in dense stellar clusters \citep[e.g.,][]{Kremer2020,Rodriguez2021, Zevin2021, Wong2021}.

\begin{table*}
	\centering
    \renewcommand{\arraystretch}{1}
    \tabcolsep=8.7pt
	\caption{Summary of simulations in \citet{Kremer2022} to be analyzed in detail in this study. In columns 2-5, we list initial conditions for the simulations. In columns 6-8, we list the total mass bound to the black hole, the final stellar mass, and the total mass unbound from the system after the first pericentre passage. In column 9, we list the orbital period of the partially disrupted star to return to pericentre (in cases where relevant). In columns 10, we describe the outcome of each simulation. The three simulations marked with an asterisk are run beyond the first passage until the star is disrupted fully.}
	\label{table:sims}
	\begin{tabular}{l|cc|cc|ccc|c|l}
\hline
\hline
\multicolumn{1}{l}{Model} &
\multicolumn{1}{c}{$M_{\rm{bh}}$} &
\multicolumn{1}{c}{$M_{\star,i}$} &
\multicolumn{1}{c}{$r_p/R_{\star}$} &
\multicolumn{1}{c}{$r_p/r_T$} &
\multicolumn{1}{c}{$M_{\rm{bound,bh}}$} &
\multicolumn{1}{c}{$M_{\star,f}$} &
\multicolumn{1}{c}{$M_{\rm{ej}}$} &
\multicolumn{1}{c}{$P_{\rm{orb}}$} &
\multicolumn{1}{c}{Outcome} \\
\multicolumn{1}{c}{} &
\multicolumn{1}{c}{($M_{\odot}$)} &
\multicolumn{1}{c}{($M_{\odot}$)} &
\multicolumn{1}{c}{} &
\multicolumn{1}{c}{} &
\multicolumn{1}{c}{($M_{\odot}$)} &
\multicolumn{1}{c}{($M_{\odot}$)} &
\multicolumn{1}{c}{($M_{\odot}$)} &
\multicolumn{1}{c}{(days)} &
\multicolumn{1}{c}{} \\
\multicolumn{1}{c}{(1)} &
\multicolumn{1}{c}{(2)} &
\multicolumn{1}{c}{(3)} &
\multicolumn{1}{c}{(4)} &
\multicolumn{1}{c}{(5)} &
\multicolumn{1}{c}{(6)} &
\multicolumn{1}{c}{(7)} &
\multicolumn{1}{c}{(8)} &
\multicolumn{1}{c}{(9)} &
\multicolumn{1}{c}{(10)} \\
\hline
1 & 10 & 0.5 & 0.70 & 0.26 & 0.299    & 0.000   & 0.201  & N/A  & Full disruption   \\ 
2 & 10 & 0.5 & 1.00 & 0.37 & 0.322    & 0.000   & 0.178   & N/A  & Full disruption    \\ 
3 & 10 & 0.5 & 1.10 & 0.41 & 0.322    & 0.000   & 0.178   & N/A  & Full disruption    \\ 
4 & 10 & 0.5 & 2.04 & 0.75 & 0.257    & 0.123   & 0.120   & N/A  & Partial disruption; stellar remnant unbound    \\ 
5 & 10 & 0.5 & 2.71 & 1.00 & 0.232    & 0.202   & 0.066   & N/A  & Partial disruption; stellar remnant unbound    \\ 
6 & 10 & 0.5 & 3.39 & 1.25 & 0.143    & 0.318   & 0.039   & N/A  & Partial disruption; stellar remnant unbound    \\ 
7 & 10 & 0.5 & 4.07 & 1.50 & 0.064    & 0.426   & 0.010   & N/A & Partial disruption; stellar remnant unbound     \\ 
8 & 10 & 0.5 & 4.48 & 1.65 & 0.031    & 0.467   & 0.002    & 303.0  & Partial disruption; stellar remnant bound    \\ 
9 & 10 & 0.5 & 4.75 & 1.75 & 0.017    & 0.483   & 0.0004  & 171.0  & Partial disruption; stellar remnant bound    \\ 
\hline 
10$^\star$ & 10 & 2   & 1.71 & 1.00 & 0.110    & 1.870   & 0.020     & 13.9  & Partial disruption; stellar remnant bound   \\
& & & & & & & & & (5 passages to full disruption) \\
\hline
11$^\star$ & 10 & 5   & 1.26 & 1.00 & 0.214    & 4.728   & 0.058    & 4.3  & Partial disruption; stellar remnant bound   \\ 
& & & & & & & & & (3 passages to full disruption) \\
\hline
12$^\star$ & 10 & 10  & 1.00 & 1.00 & 0.210    & 9.643   & 0.147     & 2.5  & Partial disruption; stellar remnant bound   \\ 
 & & & & & & & & & (3 passages to full disruption) \\
		\hline
    \hline
	\end{tabular}
\end{table*}

A third implication of the presence of black holes in stellar clusters is the occurrence of tidal disruption events (TDEs) where a black hole passes sufficiently close to a star \citep[via a single--single encounter or during a resonant binary-mediated encounter; e.g.,][]{Fregeau2004} to strip the star's outer layers or potentially disrupt the star entirely. A number of recent studies \citep[e.g.,][]{Perets2016,kremer2019tidal,Lopez2019,Wang2021,Kremer2022,Ryu2022,Kiroglu2022, Xin2023} have investigated these stellar-mass TDEs (sometimes referred to as ``micro-TDEs''). These events may lead to bright electromagnetic transient events -- for example X-ray/gamma-ray transients \citep[e.g.,][]{Perets2016} or optical/UV transients associated with reprocessing by disk wind outflows \citep[e.g.,][]{kremer2019tidal} plausibly similar to some of the fast-evolving optical transients observed to date \citep[e.g.,][]{Kremer2021_fbot}. Additionally, in the case of significant mass growth via accretion, these TDEs may imprint themselves onto the underlying black hole mass \citep[e.g.,][]{Giersz2015} and spin \citep[e.g.,][]{Lopez2019} distributions.

In \citet{Kremer2022}, we presented a suite of smoothed-particle hydrodynamics (SPH) simulations that explored the hydrodynamic outcome of these TDEs for a range in stellar masses, black hole masses, and penetration factors (the ratio of pericentre distance to the star's tidal disruption radius). We present a summary of key simulations from \citet{Kremer2022} in Table~\ref{table:sims}. As discussed in the previous paper, for a standard \citet{Kroupa2001}-like mass function, disruptions of low-mass stars ($M_{\star}\approx0.5\,M_{\odot}$) are most common. Simulations 1-9 in the table show such encounters for a variety of penetration factors. Simulations 10, 11, and 12 are examples of disruptions of more massive stars which lead to tidal capture and repeated passages.

In this study, we compute mass fallback rates following tidal disruption through post-processing analysis of the simulations of \citet{Kremer2022}. We then build a semi-analytic model for the formation and subsequent evolution of the accretion disk formed around the black hole following mass fallback. Our disk model follows the basic framework of \citet{kremer2019tidal}, but with two key differences: First, we supply directly the mass fallback rates computed from SPH simulations. Second, by leveraging the full suite of SPH simulations that cover a range of stellar masses and pericentre distances, we explore disk formation and evolution across a wide range of possible scenarios. With the disk evolution in hand, we then compute light curves across a number of different frequency bands, predicting features that in principle can be tested observationally.

This paper is organized as follows. In Section~\ref{sec:fallback}, we describe our method for computing mass fallback rates and present results. In Section~\ref{sec:disk_evol}, we describe our method for computing the disk evolution and discuss key features across various TDE scenarios (varying stellar mass and pericentre distance). In Section~\ref{sec:lightcurves}, we present light curve models computed from our disk simulations and compare to a number of observed transients in the literature. In Sections~\ref{sec:Xray} and \ref{sec:radio} we discuss prospects for producing X-ray and radio counterparts, respectively. In Section~\ref{sec:offsets} we compare the host galaxy offsets expected for these TDEs with other observed transient classes. We summarize and conclude in Section~\ref{sec:conclusion}.

\section{Mass fallback rate and disk formation}
\label{sec:fallback}

The typical method to compute fallback rate of material onto the central object following a tidal disruption is the so-called ``frozen-in'' approximation \citep[e.g.,][]{Rees1988,Ulmer1999}. In this scenario, the entire stellar mass is assumed to move with the centre of mass at the tidal radius and after disruption, the debris elements are assumed to follow independent Keplerian orbits \citep{Lodato2009}. Then the fallback rate can be written as

\begin{equation}
    \frac{dM}{dt} = \frac{dM}{dE}\frac{dE}{dt} = \frac{1}{3} \Big(2 \pi G M_{\rm bh} \Big)^{2/3} \frac{dM}{dE} t^{-5/3},,
\end{equation}
where the specific binding energy of each mass element in the disrupted stream is given by $E=G M_{\rm bh}/(2 a)$ with $a$ related to $t$, the orbital period to return to pericentre, via Kepler's third law. The frozen-in method is well-suited for the canonical TDE limit where the star is disrupted fully. In this case, a relatively flat $dM/dE$ is expected at late times, thus yielding the canonical $dM/dt \propto t^{-5/3}$ relation \citep[e.g.,][]{Rees1988} for TDEs. However, in the case of partial disruptions, the orbits of debris elements are no longer Keplerian due to the gravitational influence of the partially-stripped stellar remnant. For partial disruptions, the fallback rate can be significantly steeper than the classic $t^{-5/3}$ scaling \citep[e.g.,][]{GuillochonRamirezRuiz2013}. Using an analytic model analogous to the impulse model of \citet{Lodato2009} for full disruptions, \citet{CoughlinNixon2019} demonstrated that the fallback rate for partial TDEs is expected to scale roughly as $t^{-9/4}$, independent of the mass of the stellar core that survives the disruption. As discussed in \citet{Wang2021}, the role of the gravitational influence of the partially-stripped remnant on the fallback rate is especially important for stellar-mass black hole TDEs, where the mass of the stripped star is comparable to the black hole mass. For SPH simulations similar to those of \citet{Kremer2022} that covered a range of black hole masses (with fixed stellar mass and penetration factor), \citet{Wang2021} found the fallback rate varies from $t^{-5/3}$ to $t^{-9/4}$.

Here we compute fallback rates from our simulations in \citet{Kremer2022}, which explore a wide range in stellar masses, $M_\star \in [0.5-10\,M_{\odot}]$, and penetraction factors $r_p/r_T \in [0,2]$, where $r_p$ is the pericentre distance for the initial black hole--star orbit and $r_T$ is the star's tidal disruption radius defined here in the typical way as

\begin{equation}
    \label{eq:r_TDE}
    r_{T} = \Bigg( \frac{M_{\rm bh}}{M_{\star}} \Bigg)^{1/3} R_{\star}
\end{equation}
where $M_\star$ and $R_\star$ are the stellar mass and radius, respectively.

\begin{figure}
    \centering
    \includegraphics[width=\columnwidth]{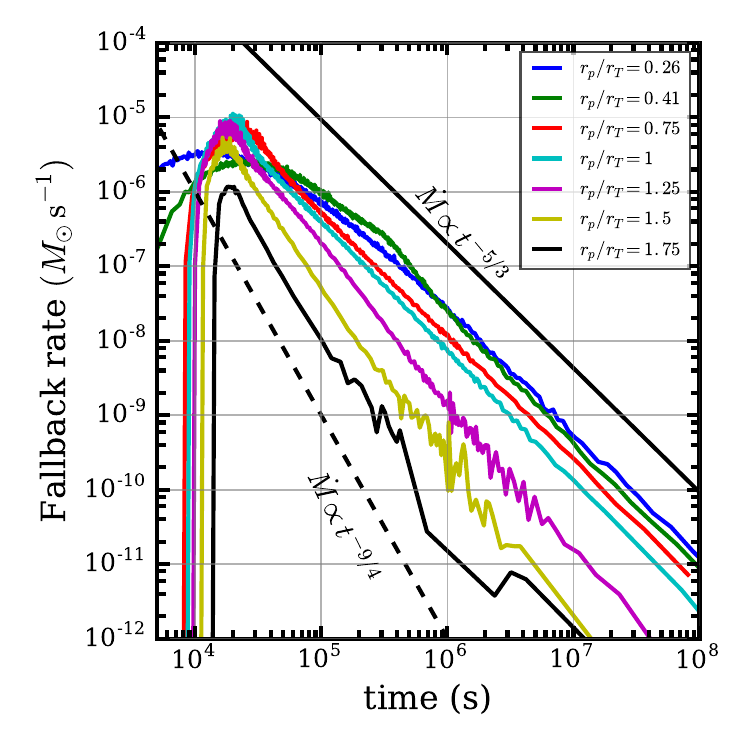}
    \caption{Mass fallback rate as computed from SPH simulations of \citet{Kremer2022} for a $0.5\,M_{\odot}$ main-sequence star interacting with a $10\,M_{\odot}$ black hole at various pericentre distances (denoted as different colors). We show here the fallback from the first pericentre passage only. For reference, we show as solid and dashed black curves the $\dot{M} \propto t^{-5/3}$ and $\propto t^{-9/4}$ fallback scalings expected for full \citep[e.g.,][]{Rees1988} and partial \citep[e.g.,][]{CoughlinNixon2019} disruptions, respectively.}
    \label{fig:fallback}
\end{figure}

\begin{figure*}
    \centering
    \includegraphics[width=\linewidth]{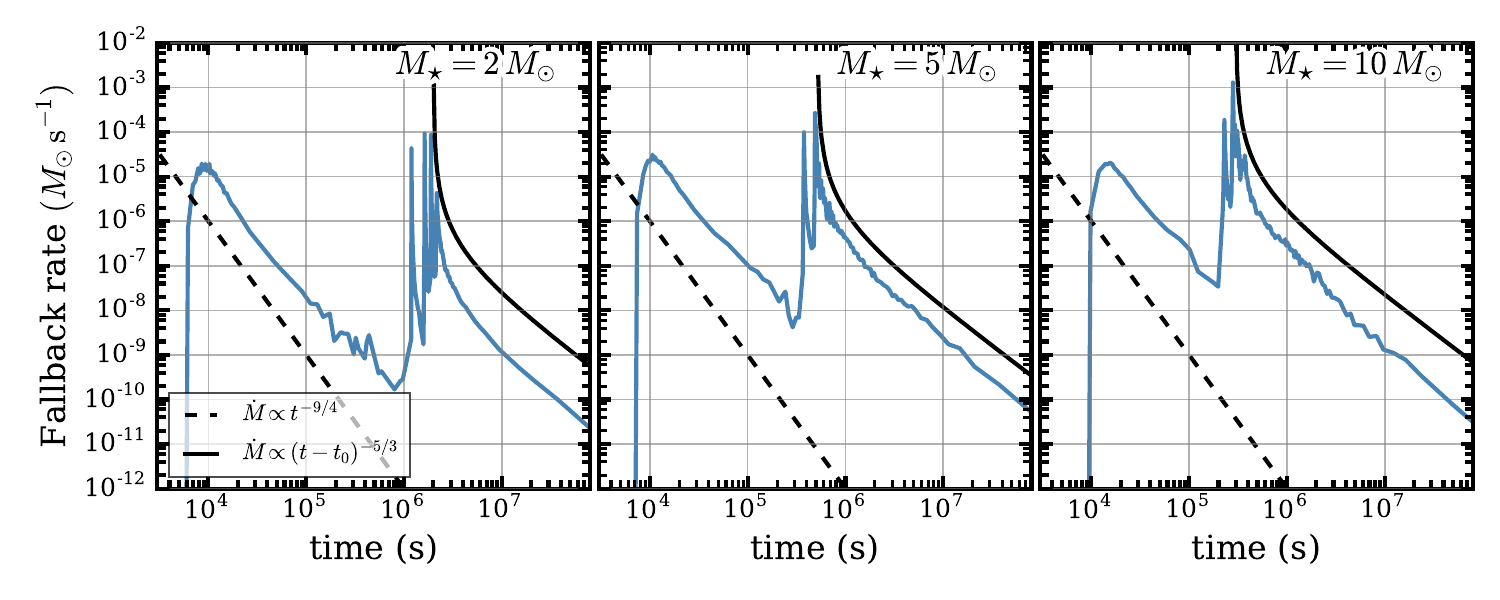}
    \caption{From left to right, mass fallback rates for main-sequence stars of masses 2, 5, and 10$\,M_{\odot}$ (simulations 10-12). In all cases the star interacts with a $10\,M_{\odot}$ black holes at initial pericentre distance $r_p=r_T$. The star is partially disrupted upon the first passage and then undergoes 3-5 additional passages before being fully disrupted. The fallback rate following the initial partial disruption follows roughly a $t^{-9/4}$ scaling, while the final full disruption follows roughly a $(t-t_0)^{-5/3}$ scaling; here we include an offset $t_0$ to account for time between first and final passages (see Figure~\ref{fig:multi_passage}).}
    \label{fig:fallback_multi}
\end{figure*}

As described in \citet{Perets2016,Wang2021}, the gravitational force due to the stripped stellar core as well as subsequent hydrodynamics (e.g., SPH particle collisions) can cause the orbits of SPH particles around the black hole to deviate from Keplerian orbits. To account for the potential deviation of debris elements from purely Keplerian orbits, we use a modified version of the frozen-in model similar to that implemented in \citet{Perets2016} when calculating the fallback rate. Unlike the standard frozen-in model, our method for determining particle return times uses multiple time snapshots from the simulations with preference given to return times calculated once a particle has withdrawn sufficiently from the star and its orbit is better described by ballistic motion around the black hole. One advantage of this approach (e.g., as opposed to a direct determination of particle return times within some characteristic radius) is it enables calculation of fallback rates well after the final time of the simulation.

Our method for determining the fallback rate consists of two main steps. First, we estimate when each SPH particle in the simulation would return to the black hole. For this, we loop forward in time through all simulation snapshots, stored at an interval of $0.25\,G^{-1/2}M_\odot^{-1/2}R_\odot^{3/2}\approx 400$\,s for the simulations in this paper. For snapshots after the first periapsis passage, we determine which particles have been stripped from the star using the same technique as in \citet{Kremer2022}. We then calculate the specific energy of each stripped particle as the sum of its specific kinetic energy (relative to the black hole) and gravitational potential energy (due to the mass of only the black hole). If this specific energy is negative (corresponding to a particle that could return to the black hole), we calculate a semimajor axis and an orbital period in the Kepler two-body approximation. If the specific energy of this particle is still negative in a subsequent snapshot, and if less than half of an orbital period has elapsed since the periapsis passage that stripped the particle, then the orbital period of the particle is recalculated and stored. Once half of an orbital period has elapsed, the stored period for that particle is ``frozen'', and its return time is determined as the sum of that period and the time of the last periapsis passage.  Should the specific energy of a stripped particle switch from negative to positive before half of an orbital period, or should the particle be reclassified as being bound to the star, then we assume that particle is not returning to the black hole; such situations can arise for particles that briefly have negative specific energy near periapsis but ultimately stay bound to the star or are ejected from the system completely.

Second, we sort the returning mass into time bins. We bin the mass according to a small initial bin width (1\,s) and then merge adjacent bins until a specified minimum number of particles per bin (200) is achieved. To avoid unreasonably large bin widths at late times, we remove the minimum particle requirement after $t=2\times 10^6$\,s (measured relative to the first periapsis) if the bin width exceeds ${\rm min}[2\times 10^{-8}\,{\rm s}^{-1} t^2,10^{11}{\rm \,s}]$. The resulting time and mass values are then used to calculate the fallback rate over time by simply dividing the total mass in each bin by the bin width. While the details of the binning procedure can affect the smoothness of the fallback rate functions, the disk evolution and resulting light curves of our model are not sensitive to such variations.

In Figure~\ref{fig:fallback} we show fallback rates for simulations 1-9 of Table~\ref{table:sims}. In these simulations, the stellar mass is fixed ($M_{\star}=0.5\,M_{\odot}$), and only the penetration factor of the encounter, $r_p/r_T$, is varied. As summarized in the table, these encounters transition from full disruption of the star ($r_p/r_T \leq 0.4$) to partial disruptions ($r_p/r_T > 0.4$) as encounters become less penetrating. As Figure~\ref{fig:fallback} shows, the transition from full to partial disruptions is accompanied by a transition in the late-time scaling of the fallback rate from $\dot{M}_{\rm fb} \propto t^{-5/3}$ to $\propto t^{-9/4}$, reproducing well the results of previous studies \citep[e.g.,][]{GuillochonRamirezRuiz2013,CoughlinNixon2019,Wang2021}.

As discussed in \citet{Kremer2022}, as the black hole to star mass ratio approaches unity, partial disruption and tidal capture of the stripped core by the black hole becomes increasingly likely. This case results in additional pericentre passages until ultimately the star is disrupted completely. Here we consider three SPH simulations of this type (see Table~\ref{table:sims}): stellar masses of $2,5,10\,M_{\odot}$ and penetration factors of $r_p/r_T=1$.
In Figure~\ref{fig:fallback_multi}, we show the fallback rate of these three simulations all the way to full disruption. As shown, in all cases the first pericentre passage follows the $t^{-9/4}$ scaling expected for partial disruptions while the final passage in which the star is disrupted fully exhibits the $t^{-5/3}$ scaling expected for full disruptions.

\section{Radiation hydrodynamics of super-Eddington accretion disks}
\label{sec:disk_evol}

By allowing the disk of material around the black hole to grow following the mass fallback rates computed in the previous section, we now introduce a method to compute the long-term evolution of the accretion disks formed.

\subsection{Disk mass and radius evolution}

Following \citet{Metzger2008} and other previous studies, we approximate the disk mass distribution as a single ring located at radius $R_d$ where the surface density distribution of the full disk peaks. We then calculate the time evolution of this ring as a proxy for the bulk properties of the disk. The time evolution of the disk is determined by conservation of mass

\begin{equation}
    \frac{d}{dt} \Big(A\pi \Sigma R_d^2 \Big) = -\dot{M_d}
\end{equation}
and conservation of angular momentum

\begin{equation}
    \frac{d}{dt} \Bigg[ B \Big(GM_{\rm bh} R_d\Big)^{1/2} \pi \Sigma R_d^2\Bigg] = \dot{J}.
\end{equation}

Here $\Sigma$ is the surface density of the disk, $M_d$ is the total disk mass, and $M_{\rm bh}$ is the mass of the central black hole. $A$ and $B$ are factors of order unity that account for the difference between the total mass and angular momentum of the disk and the mass and angular momentum near $R_d$. In this case, the total angular momentum of the disk is 

\begin{equation}
    \label{eq:J_tot}
    J=\frac{B}{A}(G M_{\rm bh}\,R_d)^{1/2}\,M_d.
\end{equation}
Finally, $\dot{M_d}$ is the total mass loss rate of the disk which, as we will discuss later, includes losses through both accretion and a disk wind.

The disk evolution is determined by the following two coupled equations:

\begin{equation}
    \label{eq:dMdt}
    \dot{M_d} = -fM_d/t_v + \dot{M}_{\rm fb}
\end{equation}

\begin{equation}
    \label{eq:Jdot}
    \dot{J} = (G M_{\rm bh} r_{\rm circ})^{1/2} \dot{M}_{\rm fb} - C (G M_{\rm bh} R_d)^{1/2} \dot{M}_{\rm out}.
\end{equation}
Here $f$ is a factor of order unity analogous to $A$ and $B$\footnote{We assume here $A/B=1$ and $f=1$ throughout, however see \citet{Metzger2008} for discussion of potentially more precise values for these parameters (factor of order unity corrections).}, $r_{
\rm circ}$ is the radius at which the bound material circularizes \citep[we assume $r_{\rm circ} = 2r_p$ as discussed in][]{Kremer2022}, and $t_v$ is the viscous accretion timescale. Adopting a standard $\alpha$-prescription for the disk \citep[e.g.,][]{ShakuraSunyaev1973} we have 

\begin{figure*}
    \centering
    \includegraphics[width=\linewidth]{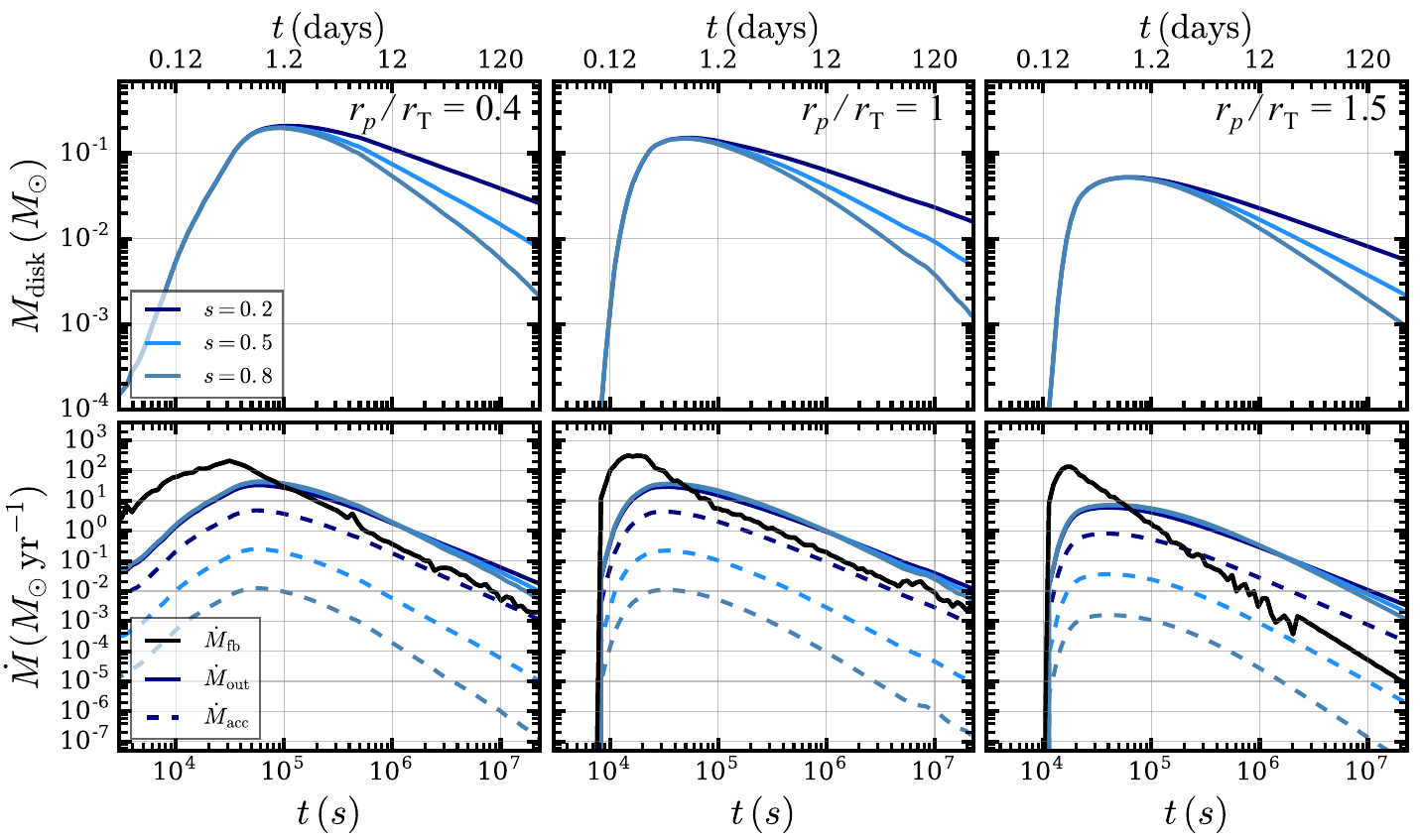}
    \caption{For simulations of $M_\star = 0.5\,M_{\odot}$ and $M_{\rm bh}=10\,M_{\odot}$, we show disk mass (top panels), mass-loss rate due to disk wind $\dot{M}_{\rm out}$ (bottom panel; solid blue curves), and mass-accretion rate onto black hole $\dot{M}_{\rm acc}$ (bottom panel; dashed blue curves) versus time after first pericentre passage. We also show the mass fallback rate as solid black curves. From left to right panels, we show show three different pericentre distances: $r_p/r_T=0.41$, $1$, and $1.5$, respectively. Different shades of blue denote different assumed values for the power-law exponent $s$ in Equation~(\ref{eq:Mdot_acc}).}
    \label{fig:masses}
\end{figure*}

\begin{multline}
    \label{eq:t_v}
    t_v = \Big[ h^2 \alpha \Omega_K(R_d) \Big]^{-1} \\ \approx 2\times10^4 \Big( \frac{h}{0.5} \Big)^{-2} \Big( \frac{\alpha}{0.1} \Big)^{-1} \Big( \frac{M_{\rm bh}}{10\,M_{\odot}} \Big)^{-1/2} \Big( \frac{R_d}{R_{\odot}} \Big)^{3/2}\,\rm{s},
\end{multline}
where $\Omega_K(R_d)=\sqrt{GM_{\rm bh}/R_d^3}$ is the Keplerian angular frequency at disk radius $R_d$ and $h=H/R_d$ where $H$ is the disk height. We assume $h=0.5$ (see \citet{Kremer2022} for discussion) and also assume $\alpha=0.1$ throughout. 

$\dot{M}_{\rm fb}$ is the growth rate of the disk due to mass fallback of the bound disruption debris. We obtain $\dot{M}_{\rm fb}(t)$ directly from our SPH simulations as described in Section~\ref{sec:fallback}. $\dot{M}_{\rm out}$ is the mass-loss rate of the disk due to disk winds. Advection-dominated disks like those expected here are likely to lose a large fraction of mass through viscously-driven outflows, which also remove angular momentum from the disk. Following previous studies \citep[e.g.,][]{Metzger2008,YuanNarayan2014,Metzger2022,Hu2022}, we assume these disk outflows cause the mass inflow rate to decrease approaching the black hole:

\begin{equation}
    \label{eq:Mdot}
    \dot{M}(r) \approx \Bigg( \frac{r}{R_d} \Bigg)^s \frac{fM_d}{t_v},
\end{equation}
\citep[e.g.,][]{BlandfordBegelman1999} where the exact value of $s \in [0,1]$ depends on the outflow model. In this case, the actual fraction of material accreted by the black hole is

\begin{equation}
    \label{eq:Mdot_acc}
    \dot{M}_{\rm acc} = \Bigg( \frac{R_{\rm acc}}{R_d} \Bigg)^s \frac{fM_d}{t_v}
\end{equation}
where we assume $R_{\rm acc} = 6GM_{\rm bh}/c^2$, the radius of the innermost stable circular orbit. Since $R_{\rm acc}/R_{\rm disk}\approx 10^{-5}$, in practice this means the overall accretion efficiency is very small. For example, for $s\approx0.5$, we expect less than $1\%$ of the total disk mass is accreted by the black hole. The total mass loss rate due to the wind outflow is

\begin{equation}
    \label{eq:Mdot_out}
    \dot{M}_{\rm out} = \Bigg[ 1 - \Bigg( \frac{R_{\rm acc}}{R_d} \Bigg)^s \Bigg] \frac{fM_d}{t_v}
\end{equation}
and the total mass loss rate of the disk due to both accretion and wind is $M_d/t_v = \dot{M}_{\rm acc} + \dot{M}_{\rm out}$. The constant $C$ (in Equation~\ref{eq:Jdot}) is determined by the torque exerted by the wind on the disk. Assuming the the outflow produces no net torque \citep[e.g.,][]{StonePringle2001}, the angular momentum losses are due only to the specific angular momentum of the outflow itself. In this case, we have \citep[e.g.,][]{Kumar2008}

\begin{equation}
    C = \frac{2s}{2s+1}.
\end{equation}

By solving Equation~(\ref{eq:J_tot}) for $R_d$ and taking the time derivative, we obtain

\begin{equation}   
    \frac{dR_d}{dt} = \frac{2 J}{G M_{\rm bh} M_d^2} \Bigg[ \dot{J} - \frac{J}{M_d} \dot{M}_d \Bigg].
\end{equation}
Combining this with Equations~(\ref{eq:Jdot}) and (\ref{eq:Mdot_out}), we obtain

\begin{equation}
    \label{eq:Rdot}
    \frac{dR_d}{dt} = \frac{2 R_d}{t_v} \Bigg[ 1 - C \Bigg( 1 - \Big[ \frac{R_{\rm acc}}{R_d} \Big]^s \Bigg) + \Bigg( \sqrt{\frac{r_c}{R_d}} - 1\Bigg) \frac{\dot{M}_{\rm fb} t_v}{M_d} \Bigg].
\end{equation}
By numerically solving Equations~(\ref{eq:dMdt}) and (\ref{eq:Rdot}), we can compute $M_d(t)$ and $R_d(t)$, and then use these solutions along with Equations~(\ref{eq:Mdot_acc}) and (\ref{eq:Mdot_out}) to also compute $\dot{M}_{\rm acc}(t)$ and $\dot{M}_{\rm out}(t)$.

\begin{figure*}
    \centering
    \includegraphics[width=\linewidth]{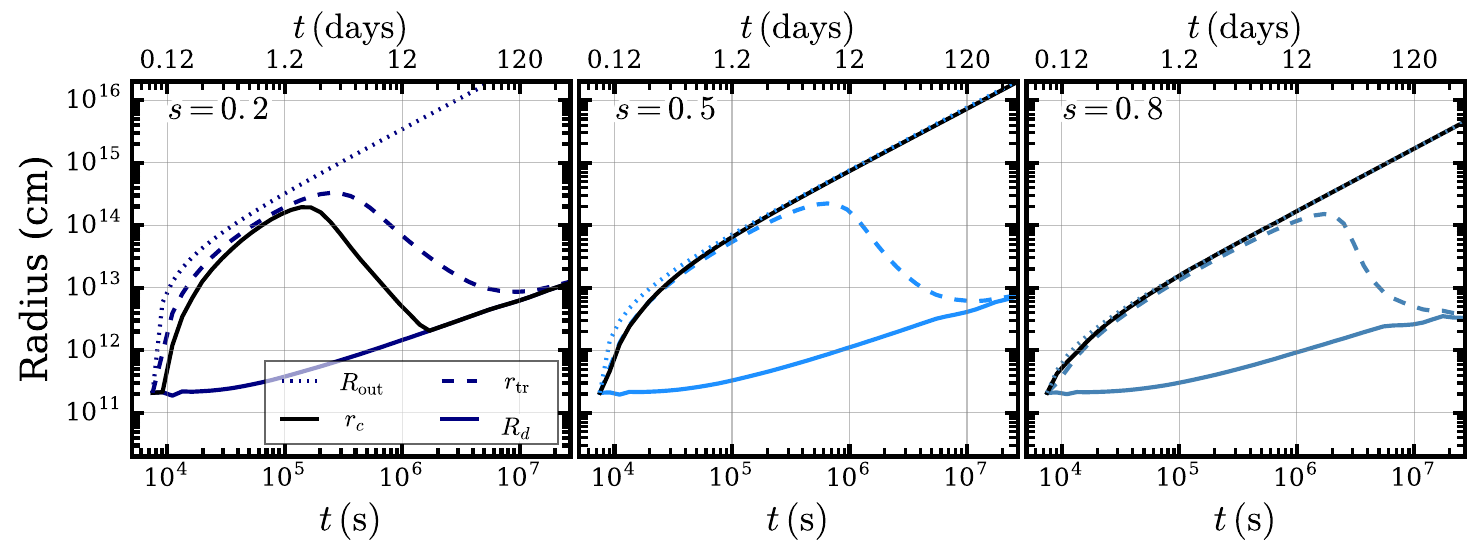}
    \caption{Disk radius $R_d$ (solid blue), trapping radius $r_{\rm tr}$ (dashed blue), color radius $r_c$ (solid black), and outer radius of disk wind $R_{\rm out}$ (dotted blue) versus time for the post-disruption evolution of encounter with $M_\star=0.5\,M_{\odot}$, $M_{\rm bh}=10\,M_{\odot}$, and $r_p/r_T=1$. From left to right, we show the evolution for different values of the power-law exponent $s$ in Equation~(\ref{eq:Mdot_acc}). For $s=0.8$, the disk becomes fallback limited after roughly $10^7\,$s, and as a result, the disk radius begins to decrease.}
    \label{fig:radius}
\end{figure*}

In Figure~\ref{fig:masses}, we show disk mass and $\dot{M}$ versus time for simulations 3, 5, and 7 in Table~\ref{table:sims} for a few different values of $s$. In the bottom panels, $\dot{M}_{\rm fb}$ is shown as solid black curves, $\dot{M}_{\rm out}$ as solid blue curves, and $\dot{M}_{\rm acc}$ as dashed blue. In Figure~\ref{fig:radius} we show the disk radius versus time for simulation 5 under a few assuptions for $s$. Figure~\ref{fig:radius} also shows a few additional radius values, discussed in the following subsection.

From Equation~(\ref{eq:Mdot}), we see the total disk evolution has two components, a growth component $\dot{M}_{\rm fb}$ and a loss component $M_d/t_v$. At early times $t \lesssim10^5\,$s, when $M_d$ is small, $\dot{M}_{\rm fb} > M_d/t_v$ and the disk grows in mass. Here, the evolution is fallback limited. Once a significant amount of bound material has fallen back to pericentre, $t \gtrsim 10^5\,$s, the evolution becomes dominated by the viscous accretion (marked by the point in time where black curves and solid blue curves cross in Figure~\ref{fig:masses}).

As discussed in Section~\ref{sec:fallback}, at late times, $\dot{M}_{\rm fb}$ lies between $\propto t^{-5/3}$ and $\propto t^{-9/4}$. Meanwhile, at late times the disk mass loss component $M_d/t_v$ goes as $t^{-(2s+4)/3}$ \citep[e.g.,][]{Metzger2008}. Thus, for fully-disruptive TDEs with $\dot{M}_{\rm fb}\propto t^{-5/3}$ (e.g., simulations 1-3; see Figure~\ref{fig:fallback}), for the case of $s > 0.5$, the disk is draining faster than it is growing so eventually the disk will become fallback limited once again. In practice, this occurs after sufficiently long time scales \citep[$t \gtrsim 10^8\,$s) that the super-Eddington disk model adopted here likely breaks down; see][]{kremer2019tidal}. For $s < 0.5$, the disk remains viscously limited at all times. For partial TDEs with fallback rates scaling as $t^{-9/4}$, the disk evolution remains viscously limited at all times, regardless of $s$. This conclusion is slightly different from that of \citet{Perets2016}, which argued that the evolution becomes fallback limited after roughly a few viscous times, roughly $10^5\,$s, in all cases. This difference arises because \citet{Perets2016} did not account for viscous spreading of the disk which regulates the disk mass loss.

\subsection{Accretion engine power and wind reprocessing}
\label{sec:reprocessing}

When a transient powered by an underlying energy source is embedded within a dense environment, the underlying powering source may be reprocessed \citep[e.g.,][]{Metzger2008, StrubbeQuataert2009, MargalitMetzger2016, kremer2019tidal,PiroLu2020,Tsuna2021,Calderon2021}. For post-TDE disks, the underlying energy source is the accretion power onto the black hole:

\begin{equation}
    \label{eq:L_d}
    L_{\rm acc} = \eta \dot{M}_{\rm acc} c^2,
\end{equation}
where $\dot{M}_{\rm acc}$ is given by Equation~(\ref{eq:Mdot_acc}) and $\eta$ is an efficiency factor. GRMHD simulations of super-Eddington accretion flows \citep[e.g.,][]{SadowskiNarayan2015,SadowskiNarayan2016} found $\eta\sim 0.01-0.1$ for accretion rates up to a few hundred times the Eddington limit. In these simulations, most of the accretion power is carried in fast outflows launched from small disk radii, enabling the Eddington limit to be exceeded dramatically. Admittedly, the peak accretion rates predicted in our simulations are $10^3-10^4$ times larger than those studied in these references. Nonetheless, we adopt $\eta\sim 0.1$ as our fiducial value \citep[for further discussion of the choice of $\eta$ in similar contexts, see][]{Metzger2022}.

Meanwhile the dense surrounding environment is supplied by the disk wind. The majority of the disk mass is expected to be launched from $R_d$ with velocity comparable to the local Keplerian orbital velocity $v_{\rm K} \sim 10^3 (M_{\rm bh}/10\,M_{\odot})^{1/2}(R_d/R_{\odot})^{-1/2}\,\rm{km\,s}^{-1}$ \citep{MargalitMetzger2016} with total power $\dot{M}_{\rm out} v_{\rm K}^2/2$. A much smaller fraction of the disk wind (of mass comparable to $M_{\rm acc}$) will be ejected near $R_{\rm acc}$ at relatively high velocities ($v \gtrsim 0.1c$) which carries a combined radiative and kinetic power of roughly $L_{\rm acc}$ \citep[e.g.,][]{SadowskiNarayan2015,SadowskiNarayan2016}. Under this outflow prescription (Equation~\ref{eq:Mdot}), large radii, $r\sim R_d$, dominate the mass budget of the outflow while small radii, $r\sim R_{\rm acc}$, dominate the energy budget. 

As the fast ejecta collides with the slower ejecta and shocks, the total engine power $L_{\rm acc}$ is expected to be thermalized near $R_d$. Assuming \citep[as in][] {kremer2019tidal} that as the shocked wind expands in radius nearly all shock heating is converted into bulk kinetic energy due to adiabatic expansion, we can estimate the (time-dependent) asymptotic wind velocity as $\dot{M}_{\rm out}v_w^2 = L_{\rm acc}$ which gives

\begin{equation}
    \label{eq:vw}
    \frac{v_w}{c} = \eta^{1/2} \Bigg(\frac{\dot{M}_{\rm acc}}{\dot{M}_{\rm out}} \Bigg)^{1/2} \approx \eta^{1/2} \Bigg(\frac{R_{\rm acc}}{R_d}\Bigg)^{s/2}. 
\end{equation}
Thus as $R_d$ increases with time (see Figure~\ref{fig:radius}), $v_w$ decreases. In this case, the outer radius of the expanding wind shell at any time can be computed simply as

\begin{equation}
    \label{eq:r_out}
   R_{\rm out}(t)~=~v_w(t=0)\times~t. 
\end{equation}
We show $R_{\rm out}$ versus time as dotted curves in Figure~\ref{fig:radius}.

\begin{figure}
    \centering
    \includegraphics[width=0.9\columnwidth]{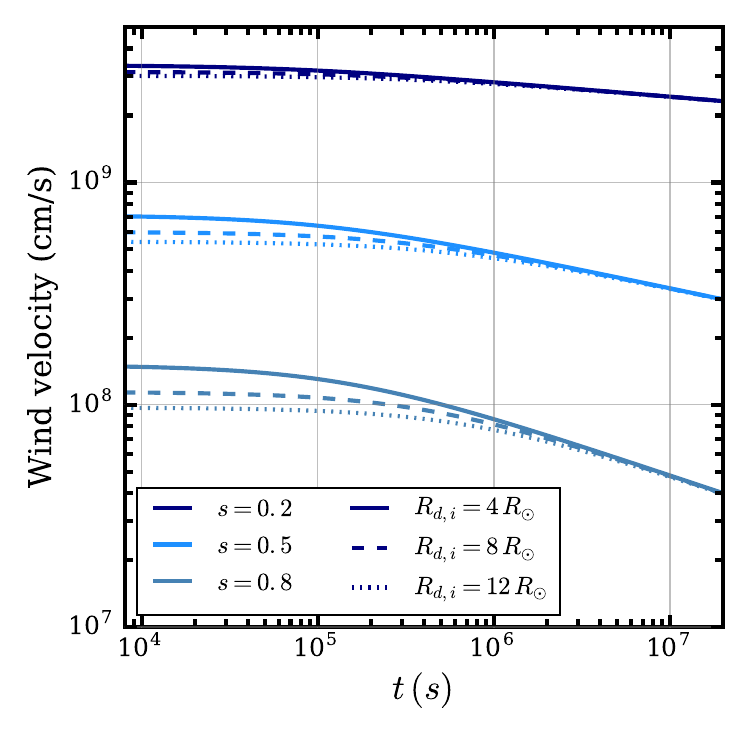}
    \caption{Asymptotic wind velocity versus time computed from Equation~\ref{eq:vw} for a few different values for $s$ and initial disk radius $R_{d,i}$ as described in the text.}
    \label{fig:vw}
\end{figure}

We show in Figure~\ref{fig:vw} the asymptotic wind velocity versus time for a few different values for $s$ and for the initial disk radius, $R_{d,i}=2r_p$. We adopt $R_{d,i}$ representative of the expected range from our SPH simulations. For $M_{\star}=0.5\,M_{\odot}$, $M_{\rm bh}=10\,M_{\odot}$, we have from Equation~(\ref{eq:r_TDE}), $r_{\rm TDE}\approx 2.7R_{\star}\approx 1.9R_{\odot}$ (for $R_{\star}=0.7R_{\odot}$). For $r_p=r_{\rm TDE}$, we predict $R_{d,i}\approx3.8R_{\odot}$. As the other extreme, consider $M_{\star}=10\,M_{\odot}$, $M_{\rm bh}=10\,M_{\odot}$, and $r_p=1.5r_{\rm TDE}$. In this case we have $r_{\rm TDE}=R_{\star}\approx4\,R_{\odot}$ and $R_{d,i}=2r_p\approx12\,R_{\odot}$. We also show $R_{d,i}=8R_{\odot}$ as an intermediate case. As shown in the figure, we expect $v_w$ values in the range $\approx0.01c-0.1c$ at all times.

Next, we compute the photon-trapping radius, $r_{\rm tr}$, defined as the radius within the wind ($r\in [R_d,R_{\rm out}]$) at which the photon diffusion time

\begin{equation}
    \label{eq:t_diff}
    t_{\rm diff}(r) = \frac{\tau (r)}{c} \frac{(R_{\rm out} - r) r}{R_{\rm out}}
\end{equation}
is equal to the dynamical time

\begin{equation}
    \label{eq:t_dyn}
    t_{\rm dyn}(r) = t - t_0(r),
\end{equation}
where $t_0(r)$ is the time at which the wind shell at current radius $r$ was originally launched (from launching radius $R_d(t_0)$ and velocity $v_w(t_0)$. In practice, $t_0(r)$ is found by solving

\begin{equation}
    \label{eq:t_0}
    r = R_d(t_0) + v_w(t_0)(t-t_0).
\end{equation}

$\tau(r)$ is the Thomson scattering optical depth of the wind outside radius $r$ which can be computed as

\begin{equation}
    \label{eq:tau}
    \tau(r) = \int_r^{R_{\rm out}} \kappa_s \rho(r^\prime) dr^\prime,
\end{equation}
where $\kappa_s=0.34\rm{cm}^2\rm{g}^{-1}$  is the opacity for electron scattering (for a solar-like composition; we do not consider the effect of varying metallicity) and where the mass density profile of the wind $\rho(r)$ is given by

\begin{equation}
    \rho(r) = \frac{\dot{M}_{\rm out}(t_0)}{4 \pi r^2 v_w(t_0)},
\end{equation}
again taking into account that $\dot{M}_{\rm out}$ and $v_w$ at a given radius are determined by their values at the time the wind was launched, $t_0$. We then compute $r_{\rm tr}$ at a given time by identifying the $r$ value that equates Equations~(\ref{eq:t_diff}) and (\ref{eq:t_dyn}), using Equations~(\ref{eq:t_0}) and (\ref{eq:tau}) to compute $t_0(r)$ and $\tau(r)$, respectively. We show $r_{\rm tr}$ versus time as dashed blue curves in Figure~\ref{fig:radius}. As shown, $r_{\rm tr}$ is reduced for smaller values of $s$, since such cases lead to lower wind mass densities which means one must go deeper into the wind to order to attain sufficiently high optical depth.

\begin{figure*}
    \centering
    \includegraphics[width=0.8\linewidth]{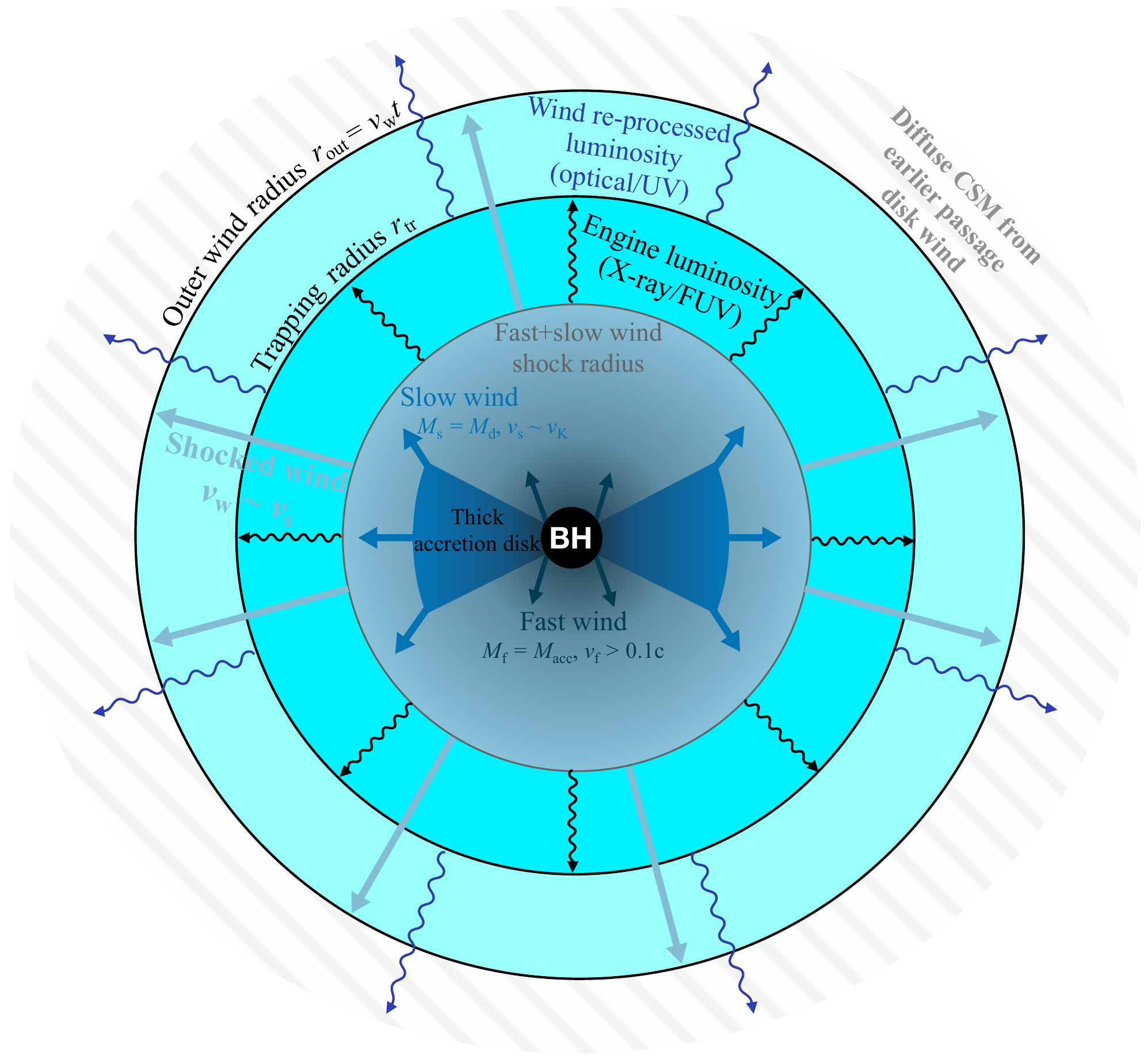}
    \caption{Cartoon illustration of the various features of our model. The thick super-Eddington disk ($r \lesssim R_d \sim 10^{11}\,$cm) launches fast outflow at $r\sim r_{\rm ISCO}$ of mass $M_f\sim M_{\rm acc}$ and velocity $v_f\gtrsim0.1c$ which collides with slower ($v_s \lesssim 0.01c$) disk wind outflow launched and shocks at $r\sim R_d$ producing engine luminosity $L_d$ (Equation~\ref{eq:L_d}). This engine power is absorbed and reprocessed by the shocked slow wind at the photon trapping radius $r_{\rm tr}$ and re-emitted (Equation~\ref{eq:L_obs}), primarily in the optical/UV. At much larger radii ($r \gg R_{\rm out} = v_w t$), the disk wind may sweep up circumstellar material from analogous disk wind launched during a (possible) earlier partial disruption, shock, and produce synchrotron emission (Section~\ref{sec:radio}).}
    \label{fig:cartoon}
\end{figure*}

With the time evolution of the photon-trapping radius in hand, we can then estimate the bolometric luminosity of the reprocessed radiation that escapes. For $r>r_{\rm tr}$, there is negligible adiabatic cooling and thus the observed bolometric luminosity remains roughly constant. As in \citet{PiroLu2020}, the observed emission is thus computed as the flux of radiation across the trapping depth:

\begin{equation}
    \label{eq:L_obs}
    L_{\rm obs}(r_{\rm tr}) = 4 \pi r_{\rm tr}^2 \mathcal{E}(r_{\rm tr}) \Bigg[v_w(t_0) - \frac{dr_{\rm tr}}{dt}\Bigg],
\end{equation}
where for $r<r_{\rm tr}$ the radiation energy density $\mathcal{E}(r)$ is computed via adiabatic expansion \citep[e.g.,][]{StrubbeQuataert2009}

\begin{equation}
    \label{eq:energy_density}
    \mathcal{E}(r) = \frac{L_{\rm acc} (1- e^{-\tau_x})}{8\pi R_d^2 v_w} \Bigg[ \frac{\rho (r)}{\rho (R_d)} \Bigg]^{4/3}.
\end{equation}
Here, $L_{\rm acc}$, $\rho(R_d) = \dot{M}_{\rm out}/(4 \pi R_d^2 v_w)$, $R_d$, and $v_w$ are all evaluated at $t=t_0$. The $dr_{\rm tr}/dt$ term in Equation~(\ref{eq:L_obs}) incorporates the effect of the changing trapping depth over time. The $(1-e^{-\tau_x})$ term accounts for the fractional amount of accretion power (primarily in X-rays) that becomes trapped within the flow. In this case, the fraction of the accretion power which escapes unabsorbed is

\begin{equation}
    \label{eq:L_x}
    L_{\rm esc} = L_{\rm acc} e^{-\tau_x},
\end{equation}
where $\tau_x$ is the optical depth computed as in Equation~(\ref{eq:tau}) above $r=R_d$ using $\kappa_x$, the opacity for absorption and thermalization of X-rays. Determining the precise value of $\kappa_x$ requires a detailed model for the ionization state of the accretion flow and ejecta, which is outside the scope of this work. Here we follow \citet{Metzger2022} and assume that the opacity for absorption and thermalization of X-rays is comparable to the opacity for electron scattering, $\kappa_x \approx \kappa_s$, reasonable for the temperature-density regimes considered here. Of course this is a simplification and further studies of the accretion physics in the inner disk are necessary to investigate these details. Qualitatively, if in fact the true value of $\kappa_x$ is higher (lower), we expect the X-rays to be trapped deeper (further out) within the ejecta outflow, effectively increasing (decreasing) the temperature and velocity of the inner shock where the fast and slow ejecta collide (see Figure~\ref{fig:cartoon}).

In general, at early times ($\lesssim$days after disruption) when $\tau_x \gg 1$, nearly all accretion power is absorbed and reprocessed. On timescales of months after disruption, the ejecta becomes optically thin and a significant fraction of accretion power begins to emerge. We discuss this further in Section~\ref{sec:Xray}.

\subsection{Temperature evolution}
\label{sec:temperature_evolution}

Next, we compute the temperature using the procedure outlined in \citet{PiroLu2020}. At any depth $r$, the temperature is dominated by radiation so $aT(r,t)^4 = \mathcal{E}(r,t)$, where $a$ is the radiation constant. For $r<r_{\rm tr}$, the radiation energy density is set by adiabatic cooling of Equation~(\ref{eq:energy_density}),  giving us

\begin{equation}
    \label{eq:T_r_below_rtr}
    T(r<r_{\rm tr},t) = [\mathcal{E}(r,t)/a]^{1/4}
\end{equation}
Above the trapping radius, the escaping luminosity is constant and the energy density and temperature are determined by flux limited diffusion

\begin{equation}
    \label{eq:T_r_above_rtr}
    L_{\rm obs}(t) = \frac{4 \pi r^2 a c}{3 \kappa_s \rho(t_0)} \frac{\partial T(r > r_{\rm tr},t)^4}{\partial r}.
\end{equation}
In this case the temperature is computed as

\begin{equation}
    \label{eq:T_profile}
    T(r>r_{\rm tr},t)^4 \approx \int_{r}^{R_{\rm out}} \frac{3 \kappa_s \rho(r^\prime,t_0) L_{\rm obs}(t)}{4 \pi {r^\prime}^2 a c} dr^\prime.
\end{equation}

\begin{figure*}
    \centering
    \includegraphics[width=\linewidth]{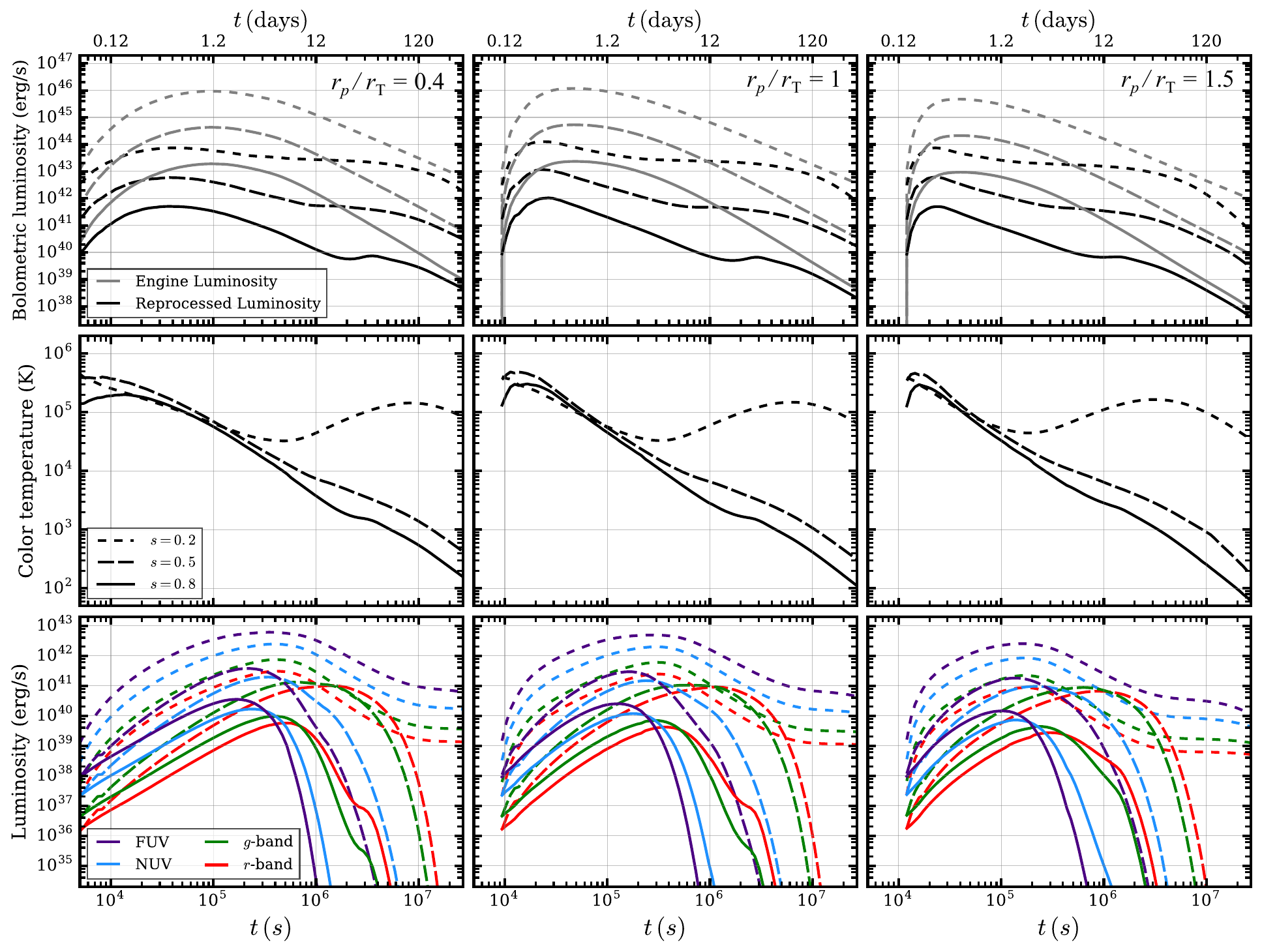}
    \caption{Luminosity and temparature evolution for $M_{\star}=0.5\,M_{\odot}$ and $M_{\rm bh}=10\,M_{\odot}$ for three different $r_p$ values: from left to right, $r_p/r_T=0.41$, $1$, $1.5$. In top panels, gray curves show the underlying engine power (Equation~\ref{eq:L_d}) and black curves show the bolometric reprocessed luminosity. Different linestyles denote different values for $s$. In middle panels, we show the color temperature evolution. Bottom panels show the reprocessed luminosity for a few different frequency bands, as defined in the text.}
    \label{fig:lum_temp}
\end{figure*}

Thermalization requires that the wind is optically thick to photon absorption. In general, the opacity for absorption, $\kappa_a$, is lower than the opacity for electron scattering, $\kappa_s$. In this case, the effective temperature of the electron scattering photosphere does not necessarily correspond to the observed color temperature. As in \citet{PiroLu2020}, we define an effective opacity (over optical/UV wavelengths)

\begin{equation}
    \label{eq:kappa_eff}
    \kappa_{\rm eff} = \sqrt{3(\kappa_a+\kappa_s)\kappa_a} \approx \sqrt{3 \kappa_s \kappa_a}
\end{equation}
(assuming $\kappa_a \ll \kappa_s$) with an associated optical depth

\begin{equation}
    \label{eq:tau_eff}
    \tau_{\rm eff}(r) = \int_r^{R_{\rm out}} \kappa_{\rm eff} \rho(r^\prime) dr^\prime .
\end{equation}
The color radius, $r_c$, is then defined as the value of $r$ that satisfies $\tau_{\rm eff}=1$.

In general, $\kappa_a$ can vary with temperature and density. Here we follow the approach of \citet{PiroLu2020} and use Kramer's opacity

\begin{equation}
    \label{eq:kappa_a}
    \kappa_a = \kappa_0 \Bigg( \frac{\rho}{\rm{g\,cm}^{-3}} \Bigg) \Bigg( \frac{T}{K} \Bigg)^{-7/2}
\end{equation}
with $\kappa_0=2\times10^{24}\,\rm{cm}^2\,\rm{g}^{-1}$. Note that for $\rho$ and $T$ values relevant here, $\kappa_a$ from Equation~(\ref{eq:kappa_a}) is much smaller than $\kappa_s$, justifying the assumption in Equation~(\ref{eq:kappa_eff}).

From Equations~(\ref{eq:T_r_below_rtr}) and (\ref{eq:T_profile}), we can compute the temperature profile $T(r,t)$ which can then be used with Equations~(\ref{eq:kappa_eff})-(\ref{eq:kappa_a}) to compute $r_c(t)$. In Figure~\ref{fig:radius}, we show the time evolution of the color radius as solid black curves.

The details of the emission that will actually be observed are determined by the observed temperature. For $r_c < r_{\rm tr}$, photons continue to adiabatically cool past $r_c$ out to $r_{\rm tr}$ due to advection. In this case (for example the $s=0.2$ panel of Figure~\ref{fig:radius}), the observed temperature is simply given by Equation~(\ref{eq:T_r_below_rtr}) evaluated at $r_{\rm tr}$. This is analogous to the assumption made in \citet{kremer2019tidal}. For $r_c > r_{\rm tr}$, photons continue to be thermalized beyond $r_{\rm tr}$ even once they are no longer advected with the flow. In this limit (for example the $s=0.5$ and $s=0.8$ cases in Figure~\ref{fig:radius}), the observed temperature is given by Equation~(\ref{eq:T_profile}) evaluated at $r=r_c(t)$:

\begin{equation}
    T_{\rm obs}(t)^4 \approx \int_{r_c(t)}^{R_{\rm out}(t)} \frac{3 \kappa_s \rho(r,t_0) L_{\rm obs}(t)}{4 \pi {r}^2 a c} dr.
\end{equation}

From the above equations, we can compute the observed bolometric luminosity of the reprocessed emission and observed temperature versus time for each SPH simulation.

In Figure~\ref{fig:cartoon}, we illustrate the key features of our model.

\section{Reprocessed emission light curves}
\label{sec:lightcurves}

\begin{figure*}
    \centering
    \includegraphics[width=\linewidth]{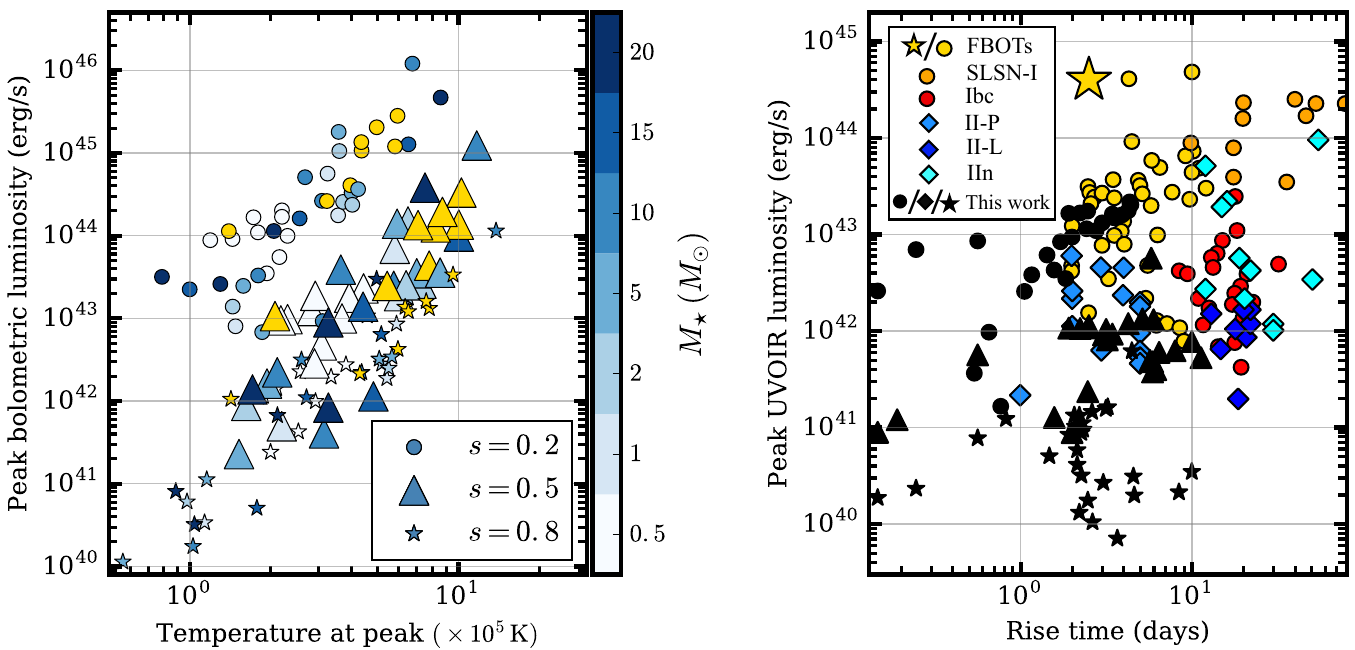}   \caption{\textit{Left panel:} Peak reprocessed luminosity versus temperature at peak following the first passage for all SPH simulations of \citet{Kremer2022} with $M_{\rm bh}=10\,M_{\odot}$ (including simulations not shown in Table~\ref{table:sims}). Different symbols denote different values of the $s$ parameter and different colors denote the different disrupted stellar masses. Blue points denote the properties after the initial pericentre passage. Yellow points indicate later pericentre passages for the three simulations of Table~\ref{table:sims} that include multiple passages. \textit{Right panel:} Peak UVOIR luminosity versus rise time to peak for all simulations compared to various observed transient classes in the literature (data taken from \citet{Margutti2019} and references therein): yellow circles are the FBOTs from \citet{Drout2014} and the yellow star is the luminous FBOT AT2018cow \citet{Margutti2019}; orange circles are superluminous SNe (SLSNe); red circles are SNe Ibc; different shaded blue diamonds are SNe II-P, II-L, and IIn. Black points denote our stellar TDEs, with different symbols denoting different $s$ values as in left panel.}
    \label{fig:lum_temp_all}
\end{figure*}

\begin{figure*}
    \centering
    \includegraphics[width=\linewidth]{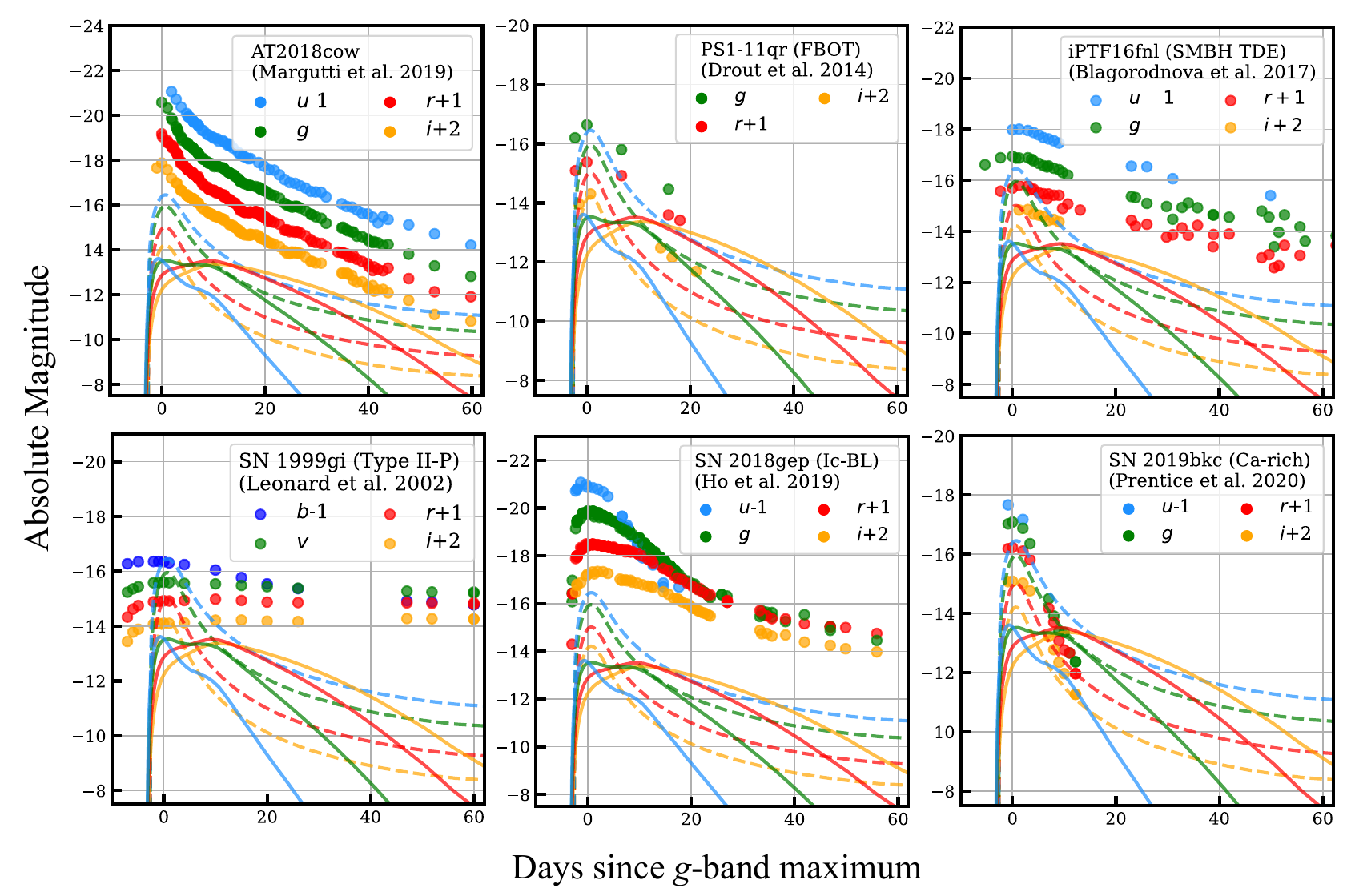}
    \caption{Optical light curves for wind-reprocessed TDE emission (showing results from simulation 5; middle panel of Figure~\ref{fig:lum_temp}) compared to a number of observed transients in the literature. Blue, green, red, and orange curves show $u$, $g$, $r$, and $i$ band emission. Dashed and solid curves show $s=0.2$ and $s=0.5$, respectively. In general, we predict these TDEs are relatively blue, fast-evolving, and dim compared to other observed transients.}
    \label{fig:lightcurves}
\end{figure*}

With our model for computing the time evolution of the disk+wind system, we can now compute the electromagnetic profiles of the reprocessed emission. In Section~\ref{sec:single_passage}, we show light curves for the case of the first pericentre passage of the SPH simulations of \citet{Kremer2022}. In Section~\ref{sec:multi_passage}, we discuss the tidal capture scenario where multiple disruptions occur. In Section~\ref{sec:rates}, we discuss the propsects for detection of these events by current/future instruments.

\subsection{Single pericentre passage}
\label{sec:single_passage}

In Figure~\ref{fig:lum_temp}, we show luminosity and color temperature versus time for the first passages of simulations 3, 5, and 7\footnote{Simulations 3 and 5 only undergo a single passage. Simulation 7 results in a bound partially stripped stellar core, so will undergo a second passage on much longer timescales.} of Table~\ref{table:sims} (from left to right). In all panels, the different line styles (solid, dashed, etc) denote different values for $s$ in Equation~(\ref{eq:Mdot_acc}).
In the top panels, the gray curves show the underlying engine luminosity powered by accretion, $L_{\rm acc}$, and black curves show the bolometric luminosity of emission reprocessed by the disk wind. In middle panels, we show the color temperature versus time. In the bottom panels, we use the reprocessed emission and color temperature to compute the observed luminosity in a few frequency bands: far-ultraviolet (FUV; defined here as $140-190\,$nm) in purple, near-ultraviolet (NUV; defined as $220-280\,$nm) in blue, $g$-band ($410-450\,$nm) in green, and $r$-band ($560-730\,$nm) in red. In order to compute the luminosity in a given wavelength band $\lambda \in [\lambda_1, \lambda_2]$ we assume blackbody emission so that

\begin{equation}
    \label{eq:L_band}
    L_{\rm band} = 4 \pi R^2\int_{\lambda_1}^{\lambda_2}  \frac{2 h c^2}{\lambda^5} \frac{1}{e^{ \frac{hc}{\lambda k_B T}}-1} d\lambda
\end{equation}
where $R=\sqrt{L_{\rm bol}/(4 \pi \sigma_{\rm SB} T_c^4)}$.\footnote{When computing in-band luminosity for a realistic detector, a frequency-dependent throughput must also be included in Equation~(\ref{eq:L_band}). We include such throughputs in Section~\ref{sec:rates}.} 

Comparison of the three columns in Figure~\ref{fig:lum_temp} shows the precise penetration factor of the tidal disruption has a relatively minor effect upon the reprocessed luminosity and temperature evolution. The accretion parameter $s$ plays a much larger role, altering the peak luminosities and time of peak by an order of magnitude or more.

At early times ($t<10^5\,$s), the color and trapping radius values are roughly similar regardless of $s$ (see Figure~\ref{fig:radius}). Thus, as seen in middle panels of Figure~\ref{fig:lum_temp}, the temperature evolution is roughly comparable at early times for different $s$ values and the luminosity values are determined mainly by the amoung of engine power injected. However, at later times, $t>10^5\,$s, the radius evolution for different $s$ values begins to diverge; for smaller $s$ values, a relatively small fraction of disk mass is launched into the wind, thus the trapping and color radii remain relatively small compared to higher $s$ values. As a result, the color temperature evolution tracks diverge which lead to distinctions in the fraction of the reprocessed luminosity emitted in different frequency bands (bottom panel of Figure~\ref{fig:lum_temp}). For $s=0.2$, the temperature remains relatively high, thus ultraviolet bands dominate at all times. For the $s=0.8$ case, where the temperature decreases most markedly, the lower frequency bands (e.g., $g$ and $r$) increasingly dominate the reprocessed luminosity output as the system evolves.

\begin{figure*}
    \centering
    \includegraphics[width=\linewidth]{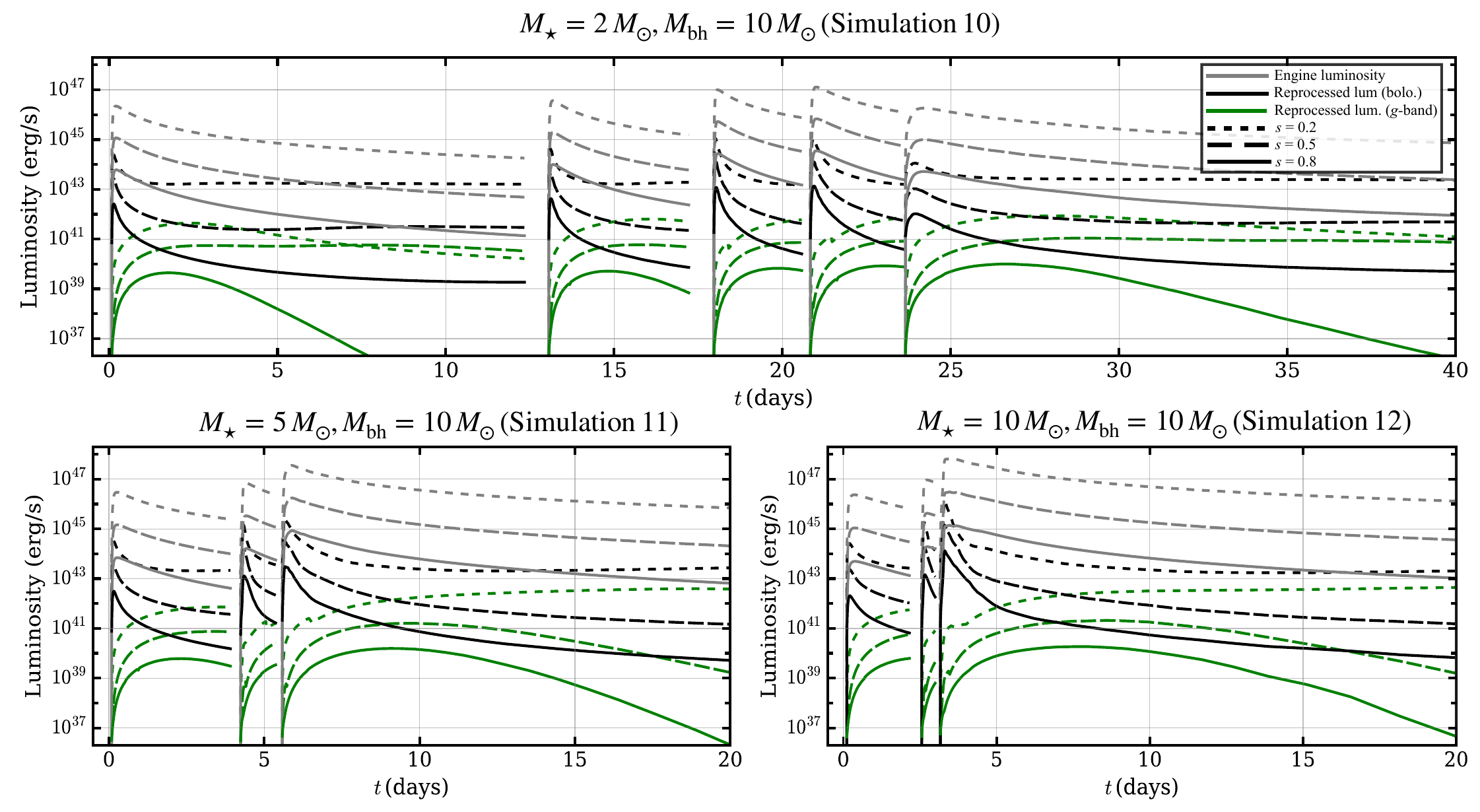}
    \caption{Light curves for each pericentre passages for the three simulations in Table~\ref{table:rates} that undergo partial disruption and tidal capture. As in Figure~\ref{fig:lum_temp}, black curves show the engine luminosity, gray curves show the bolometric reprocessed luminosity, and green show the reprocessed emission in $g$-band. As before, different linestyles denote different values for $s$.}
    \label{fig:multi_passage}
\end{figure*}

In general, Figure~\ref{fig:lum_temp} shows the peak ultraviolet emission is roughly $10\%$ of the peak bolometric value with a rise time of roughly a day to a few days (depending on $s$) following the disruption itself. Meanwhile, for the $g$ and $r$ optical bands, the peak emission is roughly $1\%$ of the peak bolometric value with relatively long rise times of a few to a few tens of days.

Figure~\ref{fig:lum_temp} shows luminosity and temperature values computed for just three specific SPH simulations. In the left-hand panel of Figure~\ref{fig:lum_temp_all}, we show in blue the peak reprocessed luminosity (bolometric) versus temperature at peak following the first passage for \textit{all} SPH simulations of \citet{Kremer2022} (including those not shown in Table~\ref{table:sims}). Different colors indicate different stellar masses. Different symbols denote different values for $s$.

For fixed $r_p/r_T$ \citep[for discussion of effect of varying $r_p/r_T$, see][]{Kremer2022}, more massive stars lead in general to brighter peak bolometric luminosities and higher temperatures. This is reasonable: more massive stars lead to more mass bound to the black hole (larger disk mass) which leads to larger engine power (peak luminosity). In turn, this leads to higher disk wind densities which lead to more compact trapping radii and higher temperatures (Equation~\ref{eq:T_profile}). Additionally, smaller $s$ values lead to larger $\dot{M}_{\rm acc}$ and thus larger engine power (peak luminosity).

In the right panel of Figure~\ref{fig:lum_temp_all}, we show peak UVOIR luminosity versus rise time to peak for all simulations (in black) compared to other stellar explosions and FBOTs in the literature (data obtain from \citet{Margutti2019} and references therein). As shown, for $s=0.5$ and especially $s=0.2$, these TDEs produce transients that most closely resemble the FBOTs \citep[e.g.,][]{Drout2014} in terms of peak luminosity and rise time, as predicted in \citet{Kremer2021_fbot}. However, the most luminous FBOTs such as AT2018cow \citep[][gold star in Figure~\ref{fig:lum_temp_all}]{Margutti2019} reach peak luminosities beyond those expected here. However, events like AT2018cow could potentially be explained in the TDE scenario for analogous disruptions involving more massive black holes \citep[$M_{\rm bh} \sim 50-100\,M_{\odot}$; e.g.,][]{Kiroglu2022} which would reach higher peak luminosities. Additionally, if the shocked fast wind region is relatively confined to the poles (as opposed to roughly isotropic as we assume here), an observer might infer a larger isotropic equivalent luminosity. This could also bring our model predictions more closely in line with the most luminous FBOTs. 

In Figure~\ref{fig:lightcurves} we compare the optical light curves from simulation 5 (middle panel of Figure~\ref{fig:lum_temp}) to a number of observed transients in the literature: AT2018cow \citep{Perley2019}; PS1-11qr, one of several fast blue optical transients (or rapidly evolving transients) from the sample in \citet{Drout2014}; iPTF16fnl \citep{Blagorodnova2017}, a relatively faint and fast TDE by a supermassive black hole; SN 1999gi \citep{Leonard2002}, a prototypical Type II-P supernova; SN 2018gep \citep{Ho2019_SN2018gep}, a recently-observed fast-rising Type Ic-BL; and SN 2019bkc \citep{Prentice2020}, a relatively fast-evolving SN Ic-like Calcium-strong transient. Compared to other observed transients, stellar-mass black hole TDEs are in general: (i) relatively fast evolving (both rise and fall time), (ii) relatively faint (especially in the case of $s\geq 0.5$ but even for our extreme case of $s=0.2$), and (iii) relatively blue, due to their high color temperatures (see Figure~\ref{fig:lum_temp_all}).

Aside from general light curve features, spectra are another critical element for classifying transients. Detailed spectral analysis is beyond the scope of this paper, however we note that disruptions of main-sequence stars by stellar black holes are in general expected to be hydrogen-rich (especially in the most common case of the disruption of low-mass stars; see Section~\ref{sec:rates}. A hydrogen-rich spectra may distinguish these TDEs quite clearly from a number of the transients shown in Figure~\ref{fig:lightcurves}. 

\subsection{Multiple passages}
\label{sec:multi_passage}

As discussed in Section~\ref{sec:fallback}, a subset of black hole--star encounters are expected to lead to partial disruption and a formation of a bound black hole--star binary that ultimately will undergo additional pericentre passages and additional disruptions.\footnote{For an animated simulation of such an encounter, visit \href{https://bpb-us-e1.wpmucdn.com/sites.northwestern.edu/dist/b/2089/files/2021/12/5_10.mov}{here}.} In Figure~\ref{fig:multi_passage}, we show the disk and luminosity evolution for each passage identified in SPH simulations 10, 11, and 12 of Table~\ref{table:sims}. We show the engine luminosity (gray), bolometric reprocessed luminosity (black), and $g$-band reprocessed luminosity (green), again for various $s$ values. The evolution for each passage is computed separately; the gaps in the light curves indicate points where each simulation is stopped and restarted.

As described in \citet{Kremer2022}, for multiple passage cases, each successive pericentre passage penetrates deeper into the star, removing more successively more mass until ultimately on the final passage, the star is disrupted entirely. Thus, (as also shown in Figure~\ref{fig:fallback_multi}), in general, the mass fallback rate increases with each pericentre passage and as a result, the engine luminosity and reprocessed bolometric luminosity generally increase with each pericentre passage. However, the orbital period decreases with each passage as energy is removed from the orbit. As the star approaches the final passages, the orbital period becomes comparable to or less than the rise time for UVOIR emission (typically $\mathcal{O}$(day); see Figure~\ref{fig:lum_temp_all}). As a result, the light curves for intermediate passages may not necessarily exhibit distinct peaks and instead may ``blur'' together and exhibit more of a ``plateau''-like profile. Of course, the final passage should exhibit a peak typical of the single passage cases described in Figure~\ref{fig:lum_temp}. 

In the left panel of Figure~\ref{fig:lum_temp_all}, we show the results of the subsequent passages for these three simulations as yellow symbols. As shown, the later passages do not exhibit noticeably distinct signatures compared to the overall population of initial pericentre passages (shown in blue) for the range in stellar mass and $r_p/r_T$ values considered.

\begin{figure}
    \centering
    \includegraphics[width=\columnwidth]{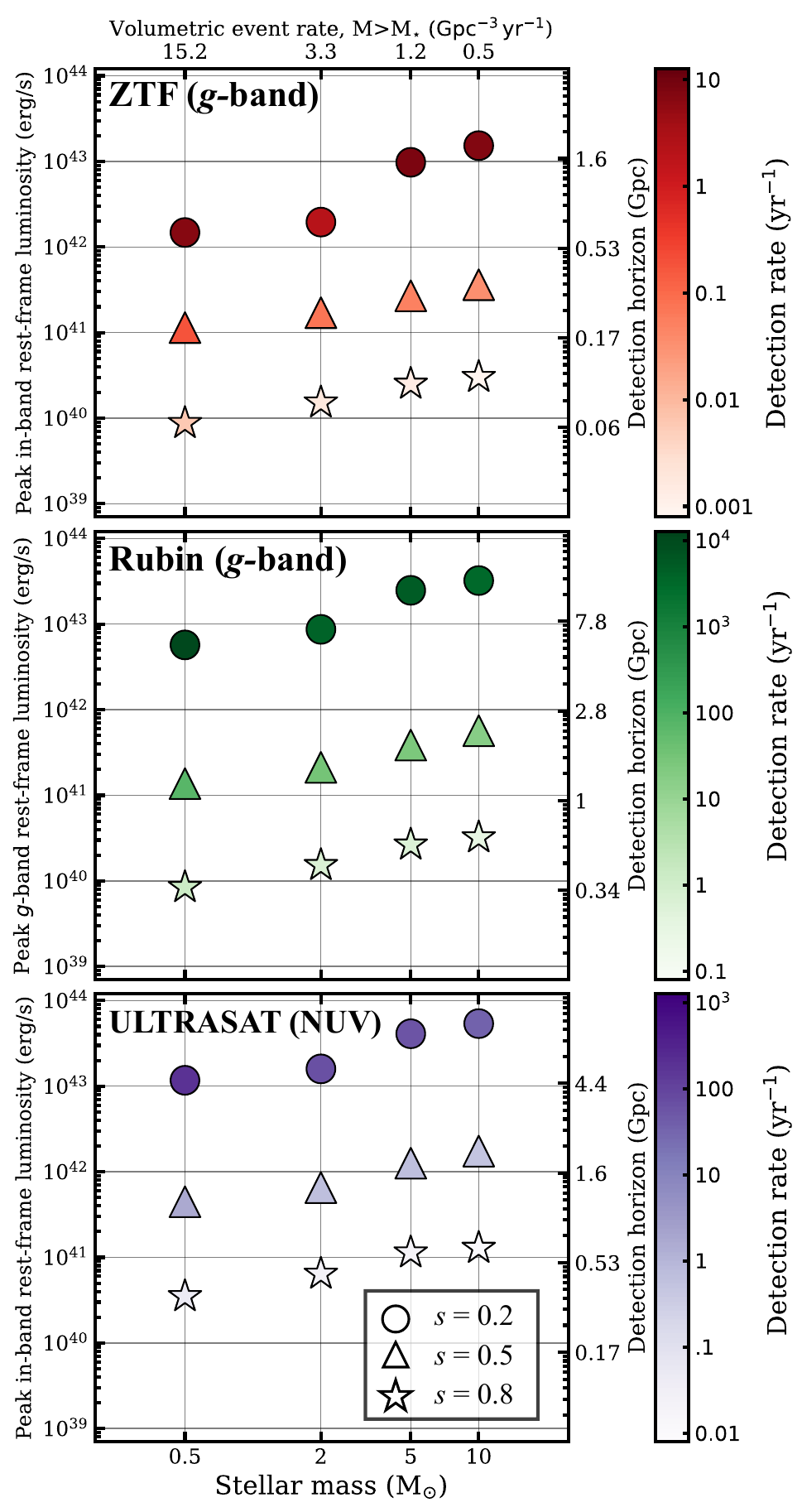}
    \caption{Detection rates for ZTF, Rubin, and ULTRASAT for TDEs of various stellar masses (lower horizontal axes). The upper horizontal axes show the total TDE event rate for stars above a given mass (from Equation~\ref{eq:rate}). The left-hand horizontal axes show the peak rest-frame luminosity in each detector's frequency band and the right-hand horizontal axes show the maximum luminosity distance an even of a given luminosity can be seen for each detector. Different symbols denote different values for $s$ in Equation~(\ref{eq:Mdot_acc}). We summarize the total detectable event rates for each detector in Table~\ref{table:rates}.}
    \label{fig:detection}
\end{figure}

\subsection{Estimates of detection rate}
\label{sec:rates}

Next we estimate the detection rates for a few different current/future instruments. We consider the $g$-bands of the Zwicky Transient Facility (ZTF) and the Vera Rubin Observatory and the NUV band of ULTRASAT. For ZTF, we adopt a limiting $g$-band magnitude of $20.8$ \citep{ZTF_2019}, for Rubin we adopt a limiting $g$-band magnitude of $25.0$ \citep{Rubin_2019}, and for ULTRASAT, we adopt a NUV limiting magnitude of 22.4 \citep{ULTRASAT_2022}.\footnote{We download frequency-dependent throughputs for each instrument from the Filter Profile Service of the Spanish Virtual Observatory project (\url{https://svo.cab.inta-csic.es}).}

As discussed in \citet{Kremer2022}, for a given range in stellar masses, the event rate can be estimated as

\begin{equation}
    \label{eq:rate}
    \Gamma \approx 9 \, \int_{m_1}^{m_2}m^{-2.08} \,dm  \,\rm{Gpc}^{-3}\,\rm{yr}^{-1},
\end{equation}
where we have assumed a \citet{Kroupa2001} mass function, a main-sequence mass-radius relation of $R\propto M^{0.6}$, and that the TDE cross section scales linearly with pericentre distance (appropriate in the gravitational-focusing regime of lower-mass star clusters) for encounters ranging from $r_p=0$ to $r_p=2r_T$. Integration of Equation~(\ref{eq:rate}) over the full mass function (e.g., from roughly $0.1-100\,M_{\odot}$), yields a total rate of roughly $100\,\rm{Gpc}^{-3}\,\rm{yr}^{-1}$, consistent with the predictions for young star clusters discussed in \citet{Kremer2021_fbot}.

Combining the event rates per stellar mass computed from Equation~(\ref{eq:rate}) with the peak in-band luminosities computed from Equation~(\ref{eq:L_band}) and detection thresholds for our selected detectors, we compute the detection rates. We show the results of this calculation in Figure~\ref{fig:detection}. The rates shown here adopt results for the cases of $r_p/r_T=1$; we assume this case is representative of all penetration factors; reasonable given, for example, Figure~\ref{fig:lum_temp}. For the $M_{\star}=0.5\,M_{\odot}$ case (simulation 5 in Table~\ref{table:sims}), we show the results after the first pericentre passage only (because the remaining stellar remnant is unbound). For the $M_{\star}=2,5,10\,M_{\odot}$ cases, we show results for the \textit{final} pericentre passage since the disk evolution of the final passage is not interrupted by later passages and, in general, the final passage is brightest.

The lower horizontal axes of Figure~\ref{fig:detection} show the mass of the disrupted star, while the upper horizontal axes show the cumulative event rate above that stellar mass (e.g., the integral in Equation~\ref{eq:rate} evaluated from $m_1=M_{\star}$ to $m_2=100\,M_{\odot}$). The left-hand vertical axes show the peak rest-frame luminosity of the various events for each detector's frequency band and the right-hand vertical axes show the maximum luminosity distance, $L_d$, at which a given luminosity could be detected for each detector's observation threshold. Detection rates (shown as different colors) are computed simply as $\mathcal{R(M_{\star})} \times (4 \pi/3 L_d^3)$, where $\mathcal{R}$ is the event rate for a given stellar mass. Different symbols denote different values of the $s$ parameter.

As shown in Figure \ref{fig:lum_temp_all}, more massive TDEs lead to brighter transients which are thus detectable out to larger distances. However, these more massive brighter events are intrinsically rarer. As Figure~\ref{fig:detection} shows, the rarer brighter events contribute roughly comparably (to within a factor of a few) to the detection rate compared to the more common less luminous events.

In Table~\ref{table:rates}, we show the total detection rates computed by dividing the full stellar mass function from $0.1-100\,M_{\odot}$ into bins centred on the stellar masses of the simulations in Table~\ref{table:sims} ($M_{\star}=0.5,2,5,10\,M_{\odot}$). We compute the TDE event rate within each mass bin using Equation~(\ref{eq:rate}) and compute the horizon distance for a given detector using the peak in-band luminosity as in Figure~\ref{fig:detection}. For old globular clusters (which have a much narrower stellar mass function at present day, from roughly $0.1-1\,M_{\odot}$), we use only the results of the $M_{\star}=0.5\,M_{\odot}$ simulation to compute the detection rate.

\begin{table*}
	\centering
    \renewcommand{\arraystretch}{1}
    \tabcolsep=8.7pt
	\caption{Detection rates for stellar-mass black hole TDEs for different cluster environments and instruments. Here we show \textit{total} detection rates integrated over all stellar masses, as described in text. In Figure~\ref{fig:detection} we show detection rates per stellar mass.}
	\label{table:rates}
	\begin{tabular}{l|l||c|c|c|c} 
\hline
\hline
\multicolumn{1}{c}{Environment} &
\multicolumn{1}{c}{Intrinsic rate} &
\multicolumn{1}{c}{$s$} &
\multicolumn{1}{c}{Rubin ($g$-band)} &
\multicolumn{1}{c}{ZTF ($g$-band)} &
\multicolumn{1}{c}{ULTRASAT (NUV)} \\
\multicolumn{1}{c}{} &
\multicolumn{1}{c}{($\rm{Gpc}^{-3}\rm{yr}^{-1}$)} &
\multicolumn{1}{c}{} &
\multicolumn{1}{c}{($\rm{yr}^{-1}$)} &
\multicolumn{1}{c}{($\rm{yr}^{-1}$)} &
\multicolumn{1}{c}{($\rm{yr}^{-1}$)} \\
\multicolumn{1}{c}{(1)} &
\multicolumn{1}{c}{(2)} &
\multicolumn{1}{c}{(3)} &
\multicolumn{1}{c}{(4)} &
\multicolumn{1}{c}{(5)} &
\multicolumn{1}{c}{(6)} \\
\hline
Globular clusters & $\sim$10 \citep{Perets2016,kremer2019tidal} & 0.2 & $6.8\times10^3$ & 4.3 & 133 \\
 & & 0.5 & $50$ & 0.1 & 1.2 \\
  & & 0.8 & $0.9$ & 0.004 & 0.03 \\
\hline
Young massive clusters & $\sim$100 \citep{Kremer2021_fbot} & 0.2 & $6.8\times10^4$ & 53.9 & $1.3\times10^3$\\
 & & 0.5 & $490$ & 1.1 & 12 \\
  & & 0.8 & $9.1$ & 0.9 & 0.3 \\
	\hline
    \hline
	\end{tabular}
\end{table*}

\section{Emergence of unabsorbed engine luminosity: high-energy counterpart}
\label{sec:Xray}

As the disk wind becomes optically thin (see Equation~\ref{eq:tau}), a significant fraction of the underlying engine power may escape unabsorbed revealing a high-energy counterpart alongside the reprocessed emission discussed in the previous section. As shown in Figure~\ref{fig:lum_temp}, the typical timescale for the disk to become optically thin is roughly months after disruption. At this time, the engine power ranges from roughly $10^{40}-10^{44}\,\rm{erg/s}$, depending on $s$. Assuming equipartition of the shock that powers the engine (see Figure~\ref{fig:cartoon})

\begin{equation}
    \rho v_w^2 = aT_{\rm sh}^4,
\end{equation}
where $\rho = \dot{M}_w/(4 \pi R_d^2 v_w )$, $v_w$ is the wind velocity from Equation~(\ref{eq:vw}), and $R_d$ is the disk radius, we can compute the temperature of the shock, $T_{\rm sh}$. At $t \sim 10^7\,$s, we find temperatures ranging from roughly $8\times10^4\,$K (for $s=0.8$) up to roughly $2\times10^5\,$K (for $s=0.2$), corresponding to peak blackbody emission in the extreme UV/soft X-rays.

\subsection{Jet formation}

Alternatively/additionally, a fraction of the engine power may escape at relatively early times. The two-zone disk model adopted here consisting of a slow and fast wind component is a simplification. In reality, a spread of velocities is expected. If a fraction of the fastest wind from the inner disk is able to pierce through the slow wind without being stalled and shocking as described in the basic picture of Figure~\ref{fig:cartoon}, then a ``jet''-like geometry is expected. In this case, the high-energy power emitted from the inner disk is absorbed by much smaller fraction of material within the jet and a much larger fraction may escape unabsorbed.

To consider this possibility, we adopt an approach similar to that of \citet{Metzger2022}. Assume mass $M_{\rm acc} \ll M_d$ is launched from the innermost disk radii near $r_{\rm ISCO}$ at constant velocity $v_f$. Assume that as this ``ultra''-fast wind interacts with the wider-angle slow disk outflows, it becomes collimated along the disk rotation axis and creates a jet-like geometry. The optical depth within the jet is 

\begin{equation}
    \label{eq:tau_jet}
    \tau_{\rm jet} \approx  \frac{M_{\rm acc} \kappa}{4 \pi f (v_f t)^2} \sim 1 \Bigg(\frac{f}{0.1} \Bigg)^{-1} \Bigg( \frac{M_{\rm acc}}{10^{-3}\,M_{\odot}} \Bigg) \Bigg( \frac{v_f}{0.5c} \Bigg)^{-2} \Bigg( \frac{t}{12\,\rm{hr}} \Bigg)^{-2}
\end{equation}
for $\kappa = 0.34\,\rm{cm}^2\,\rm{g}^{-1}$ (see Section~\ref{sec:reprocessing}). The factor $f$ is the angle subtended by the jet. The specific value of $v_f$ depends on the details of the inner disk which are outside the scope of our study. A fiducial value of $v_f \approx 0.5c$ is adopted in Equation~(\ref{eq:tau_jet}) in line with the discussion in \citet{Metzger2022} as a possible upper limit on the ejecta velocities in some of the observed luminous FBOTs. Additionally, GRMHD simulations \citep[e.g.,][]{SadowskiNarayan2015} of super-Eddington accretion disks analogous to those studied here predict maximum velocities comparable to this value. Higher (lower) values of $v_f$ would reduce (increase) the optical depth within the jet and therefore reduce (increase) the characteristic timescale $t$ for release of the reprocessed emission.

We have assumed as in \citet{Metzger2022} that the relatively high $\dot{M}_{\rm acc}$ values at early times cause large radii to dominate the optical depth integral. In this case (see Equation~\ref{eq:L_x}), a significant fraction of the engine power (emitted primarily as X-rays) from the inner disk would escape unabsorbed around a few hours to a day after disruption. The luminosity associated with this mechanism, roughly $\eta \dot{M}_{\rm acc} c^2 \sim 10^{46} (\dot{M}_{\rm acc}/10\,M_{\odot}\rm{yr}^{-1})\,\rm{erg\,s}^{-1}$ (for $\eta=0.03$), is the maximum possible luminosity expected from these TDEs \citep[see also discussion in][]{Perets2016,kremer2019tidal,Kremer2022}.

\section{Shock-powered radio emission}
\label{sec:radio}

Bright synchrotron radio and millimeter emission is a defining feature of the several luminous FBOTs including the AT2018cow, ZTF18abvkwla, and CSS161010 events. This radio emission is
consistent with self-absorbed synchrotron radiation produced from an external shock generated as the ejecta interacts with a dense external medium \citep[e.g.,][]{Ho2019,Margutti2019,Ho2020,Coppejans2020}. Following the standard framework for self-absorbed synchrotron emission from SNe \citep[e.g.,][]{Chevalier1998}, \citet{Margutti2019,Ho2020,Coppejans2020} inferred circumstellar medium (CSM) densities ranging from roughly $10-10^6\,\rm{cm}^{-3}$ for AT2018cow, ZTF18abvkwla, and CSS161010, respectively, for various observation epochs and for a range of microphysics assumptions.

Here we examine whether the TDEs described here may plausibly host sufficiently high CSM densities to power radio emission similar to that observed for these luminous FBOTs. Consider the TDE scenario involving a tidal capture and a series of repeated passages en route to full disruption of the star, as in simulations 10-12 (Figure~\ref{fig:multi_passage}). Each successive passage will result in its own disk-wind ejection episode. As the disk-wind from a later passage expands, it ultimately will collide with wind material launched during an earlier passage. This will occur after time $\Delta t$, comparable to the orbital period of the star-black hole binary following the initial passage. A simple estimate assuming homologous expansion (of course a more detailed treatment should consider more precise wind expansion scenarios) yields the following scaling for the ejecta density at the time of the wind-wind collision:

\begin{equation}
    \label{eq:ejecta_density}
    n_{\rm{ej}} \approx 10^5\,\rm{cm}^{-3} \, \Big( \frac{\textit{M}_{\textit w}}{10^{-2}\,\it{M}_{\odot}} \Big) \Big( \frac{\textit{v}_{\textit w}}{10^3\rm{km/s}} \Big)^{-3} \Big( \frac{\Delta \it{t}}{10\,\rm{yr}} \Big)^{-3}.
\end{equation}

\begin{figure}
    \centering
    \includegraphics[width=\columnwidth]{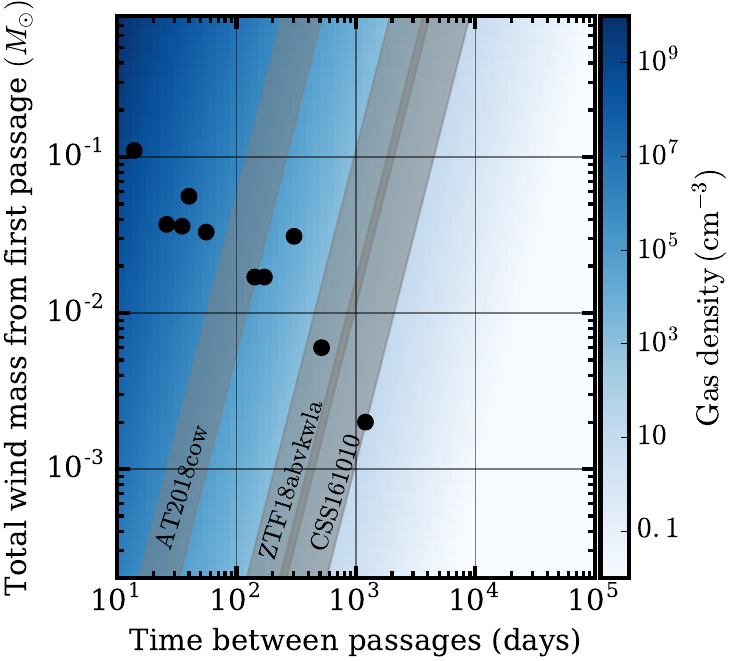}
    \caption{Average gas density predicted from wind-mass ejecta associated with first partial disruption occurring a time $\Delta t$ before the second disruption (Equation \ref{eq:ejecta_density}). Black points show the values for $M_w$ (the total mass bound to the black hole) and $\Delta t$ (the orbital period of partially disrupted stellar core+black hole binary) computed from our SPH simulations following the first partial disruption. Here we adopt $v_w=0.1c$ \citep[see][]{MargalitQuataert2021}. We show as gray bands the ranges of constant density inferred for AT2018cow \citep{Ho2019}, ZTF18abvkwla \citep{Ho2020}, and CSS161010 \citep{Coppejans2020} from radio observations $22$, $81$, and $99\,$days after each respective explosion.}
    \label{fig:ejecta}
\end{figure}

In Figure \ref{fig:ejecta}, we show the gas density expected from Equation~(\ref{eq:ejecta_density}) for various values of $M_w$ and $\Delta t$, assuming $v_w=0.1c$.  For reference, the gray bands mark ranges of constant density inferred from radio observations of AT2018cow \citep{Ho2019}, ZTF18abvkwla \citep{Ho2020}, and CSS161010 \citep{Coppejans2020} $22$, $81$, and $99$ days after explosion, respectively. The black scatter points mark the $[M_w,\Delta t]$ values following the first passage identified in all SPH simulations from \citet{Kremer2022} where the star is partially disrupted and where the stripped core becomes bound to the black hole. As shown, CSM densities comparable to the range inferred from these three FBOT events are naturally reproduced by our TDE simulations.

\citet{MargalitQuataert2021} argue $v \sim 0.1-0.5c$ are required to explain the spectra of the aforementioned FBOTs (e.g., see their Figure 2). Thus, under the assumptions made in our study, $s\lesssim 0.2$ may be required to produce the mildy-relativistic wind velocities that appear necessary for FBOT-like events (see Figure~\ref{fig:vw}). Non-relativistic velocities $v\sim0.01c$ corresponding to higher $s$ values can still produce shock-powered synchrotron radio emission, but likely with spectra more similar to that expected for radio supernovae \citep[e.g.,][]{Chevalier1998}.

\section{Host offsets}
\label{sec:offsets}

A key question is how to classify a given observed transient event as a stellar-mass black hole TDE. As shown in Figure~\ref{fig:lightcurves}, these events may exhibit unique light curve features compared to other events. Another useful distinguishing feature is the host environment. For instance, TDEs occurring in globular clusters are expected to feature large offsets from their host galaxy's centre. To test this, we compute projected offset distributions for TDEs. For the globular cluster offset distributions, we assume the distribution computed in \citet{Shen2019} by adopting S\'{e}rsic profile of index $n=2$ and integrating over an assumed halo mass function. Since the characteristic size of globular clusters \citep[roughly a few pc; e.g.,][]{Harris1996} is much smaller than the typical galactocentric offset of clusters ($\gtrsim 1\,$kpc), we assume the TDE offset distribution simply traces the cluster offset distribution. We do not consider how the TDE rate per globular cluster may vary with specific cluster properties (e.g., cluster mass, metallicity, half-light radius), which themselves may vary with host offset. We reserve for future work consideration of these details.

\begin{figure}
\centering
\includegraphics[width=0.92\columnwidth]{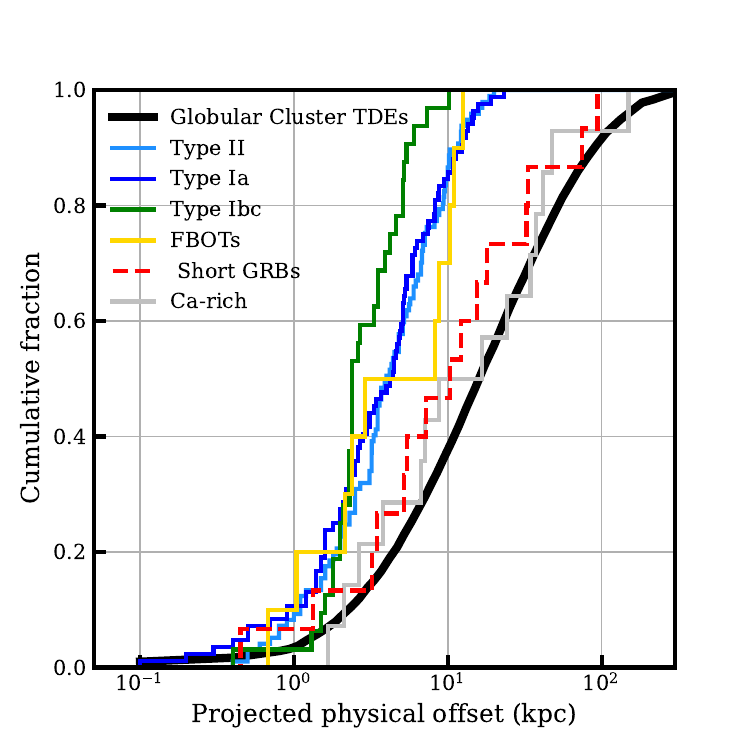}
\caption{\label{fig:offsets} Projected host offsets for TDEs occurring in globular clusters (black) in comparison to other transients in the literature: Type II SNe \citep[light blue;][]{Prieto2008}, Type Ia SNe \citep[dark blue;][]{Prieto2008}, Type Ibc SNe \citep[green;][]{Prieto2008}, FBOTs \citep[yellow;][]{Drout2014}, short GRBs \citep[dashed red;][]{FongBerger2013}, and Ca-rich transients \citep[silver;][]{Shen2019}.}
\end{figure}

In Figure~\ref{fig:offsets}, we show the distribution (cumulative fraction) of physical host offsets for a variety of transients classes (see references in figure caption) compared to our computed TDE offsets (solid black curve). As shown, TDEs occurring in globular clusters will in general have relatively large offsets compared to all observed transients, with the possible exception of the calcium-strong transients \citep[e.g.,][]{Kasliwal2012}.\footnote{Note that the intrinsic rate inferred for the calcium-strong transients, roughly $10^3\,\rm{Gpc}^{-3}\rm{yr}^{-1}$, comparable to the SN Ia rate \citep[e.g.,][]{Kasliwal2012}, is significantly higher than our predicted rate for stellar black hole TDEs.} 

As discussed in Section~\ref{sec:rates}, a high fraction of these TDEs are expected in young stellar clusters which are expected to trace more closely standard star forming environments \citep[e.g.,][]{PortegiesZwart2010}. For such TDEs, host offset is likely not a useful way to distinguish these events from, e.g., transients associated with standard core-collapse supernovae.

\section{Summary and conclusions}
\label{sec:conclusion}

Incorporating the results of the SPH simulations presented in \citet{Kremer2022}, we have explored the formation and evolution of accretion disks formed through the tidal disruption of main sequence stars by stellar-mass black holes. We then used these accretion disk models to compute light curves associated with disk-wind reprocessing. We summarize our key conclusions below:

\begin{itemize}
    \item Depending on whether the star is partially or fully disrupted, we find the mass fallback rates lie between a $t^{-5/3}$ scaling (full disruptions) and a $t^{-9/4}$ scaling (partial disruptions), consistent with predictions from previous studies of supermassive black hole TDEs \citep[e.g.,][]{GuillochonRamirezRuiz2013,CoughlinNixon2019}. However, in general the time evolution of electromagnetic signatures is determined by the accretion disk evolution, not the fallback rate.

    \item In all cases, the accretion flow in the disks formed following mass fallback is highly super-Eddington. As in \citet{kremer2019tidal,Kremer2022}, we argue a significant fraction ($\gtrsim 99\%$) of the disk mass is lost through a disk wind, with only a small amount being accreted by the black hole.

    \item The radiation from the central engine powered by the accretion of mass onto the black hole $L_{\rm acc} \sim \eta \dot{M}_{\rm acc} c^2$ is absorbed and reprocessed at radii outside the outer disk radius and re-emitted as thermal emission. Due to adiabatic expansion out to the trapping radius, the engine power is typically reduced by a factor of roughly $100$ when it emerges following reprocessing. We predict bolometric reprocessed luminosities ranging from roughly $10^{40}-10^{44}\,$erg/s. The details of the disk wind model (namely the $s$ parameter that determines the fraction of disk material launched as a wind versus accreted; Equation~\ref{eq:Mdot_acc}) are the key factor that determines the typical peak luminosity. Within the range in $s$ explored here, the luminosity can vary by factors of up to roughly $100$. Parameters such as the mass of the disrupted star and penetration factor of the encounter have a less prominent effect (factors $\lesssim 10$).
    
    \item In general, the effective temperature of this reprocessed emission is $\sim 10^5-10^6\,$K at peak luminosity. Depending again on the details of the disk wind model, the temperature can decrease to values as low as $\sim 100\,$K at late times ($t\gtrsim10^6\,$s). Our predicted temperature values imply electromagnetic signals primarily at ultraviolet/blue wavelengths. In general, these events are bluer than other observed optical transients in the literature. 

    \item For ultraviolet wavelengths representative of the near UV band of ULTRASAT, we predict typical peak luminosity of roughly $10^{41}-10^{42}\,$erg/s. For $g$-band optical wavelengths representative of ZTF and the Rubin Observatory, we predict peak luminosities of comparable values. Incorporating relevant detector sensitivities and intrinsic event rate predictions from previous studies \citep{Perets2016,kremer2019tidal,Kremer2021_fbot}, we predict detection rates ranging from roughly $10-10^5\,\rm{yr}^{-1}$ (Rubin), $1-50\,\rm{yr}^{-1}$ (ZTF), and $0.3-10^3\,\rm{yr}^{-1}$ (ULTRASAT).

    \item On longer timescales ($\gtrsim \mathcal{O}$(month) after disruption), the disk wind becomes optically thin and a significant fraction of engine luminosity can escape unabsorbed. This may lead to a late time extreme-UV/soft X-ray counterpart of peak luminosity $10^{40}-10^{44}\,$erg/s. Additionally, if a fraction of disk wind ejecta becomes collimated into a jet-like geometry, a very luminous X-ray counterpart as high as roughly $10^{46}\,$erg/s may emerge at early times (a few hours to a day after disruption).

    \item In cases where the star is partially disrupted after the first passage and the stellar core becomes bound to the black hole, successive flares likely result. Additionally, disk wind ejecta from successive passages will collide and shock with one another at large radii ($r \gg r_{\rm disk}$). Depending on density of the wind ejecta (determined by total mass of the wind and its velocity), this shock may produce a radio counterpart similar to that observed for a number of luminous fast blue optical transients such as AT2018cow through the production of self-absorbed synchrotron emission.

    \item Finally, for the subset of stellar black hole TDEs occurring in old globular clusters, we demonstrate the associated transient events will have physical offsets from their host galaxies much larger than the offsets of most other observed transients in the literature. If measured for specific transient events, such offsets may point clearly toward a stellar black hole TDE origin.
\end{itemize}

\section*{Acknowledgements}

We thank Daichi Tsuna and the anonymous referee for comments on the mansucript. Support for this work and for KK was provided by NASA through the NASA Hubble Fellowship grant HST-HF2-51510 awarded by the Space Telescope Science Institute, which is operated by the Association of Universities for Research in Astronomy, Inc., for NASA, under contract NAS5-26555. 
This research has made use of the Spanish Virtual Observatory (https://svo.cab.inta-csic.es) project funded by MCIN/AEI/10.13039/501100011033/ through grant PID2020-112949GB-I00.

\vspace{-0.7cm}
\section*{Data Availability}

The data supporting this article are available on reasonable request to the corresponding author.


\vspace{-0.7cm}

\bibliographystyle{mnras}
\bibliography{mybib} 




\bsp	
\label{lastpage}
\end{document}